\documentclass[sigconf]{acmart}
\usepackage{popets}

\usepackage{tikz}
\usepackage{graphicx}
\usetikzlibrary{positioning,arrows.meta}

\definecolor{gppibluefill}{RGB}{248,251,255}
\definecolor{gppiblueborder}{RGB}{120,155,220}
\definecolor{gppigreenfill}{RGB}{249,253,249}
\definecolor{gppigreenborder}{RGB}{126,176,120}
\definecolor{gppiorangefill}{RGB}{255,250,245}
\definecolor{gppiorangeborder}{RGB}{226,150,82}
\definecolor{gppigray}{RGB}{55,55,55}
\definecolor{gppilightgray}{RGB}{245,245,245}

\usetikzlibrary{positioning, arrows.meta}
\usepackage{xcolor}

\definecolor{blueborder}{RGB}{91,  155, 213}
\definecolor{bluefill}  {RGB}{236, 245, 255}
\definecolor{greenborder}{RGB}{84,  170,  84}
\definecolor{greenfill}  {RGB}{233, 250, 233}
\definecolor{orangeborder}{RGB}{210, 148,  60}
\definecolor{orangefill}  {RGB}{255, 246, 232}
\definecolor{textgray}    {RGB}{75,   75,  75}

\setcopyright{none}
\renewcommand\footnotetextcopyrightpermission[1]{}

\acmDOI{}
\acmISBN{}
\acmConference{}

\newcommand{\cmark}{\ding{51}}  
\newcommand{\xmark}{\ding{55}} 
\newcommand{\pmark}{$\scriptstyle\triangle$} 

\usepackage{hyphenat}
\usepackage{placeins}
\usepackage{enumitem} 

\usepackage{graphicx} 
\usepackage{booktabs}
\usepackage{array}
\usepackage{bm} 
\usepackage{colortbl}
\usepackage{makecell}
\usepackage{pifont}
\usepackage{xcolor}
\usepackage{threeparttablex}
\usepackage{rotating}

\usepackage{amsmath}

\usepackage{xcolor}

\definecolor{pastelgreen}{RGB}{119, 221, 119}
\definecolor{pastelred}{RGB}{255, 105, 97}
 
\setlist[itemize]{leftmargin=*,labelindent=0pt}
\usepackage[normalem]{ulem}
\usepackage{etoolbox}

\begin{document}

\title[Modernizing User Privacy Preference Measurement through GPPI]{Modernizing User Privacy Preference Measurement through GPPI: A GDPR-aligned Privacy Preference Item Bank}






\author{Yahya Hmaiti}
\affiliation{
  \institution{University of Central Florida}
  \city{Orlando}
  \state{Florida}
  \country{USA}
}
\email{yohan.hmaiti@ucf.edu}

\author{Mykola Maslych}
\affiliation{
  \institution{University of Central Florida}
  \city{Orlando}
  \state{Florida}
  \country{USA}
}
\email{mykola.maslych@ucf.edu}

\author{Amirpouya Ghasemaghaei}
\affiliation{
  \institution{University of Central Florida}
  \city{Orlando}
  \state{Florida}
  \country{USA}
}
\email{aghaei.ap@ucf.edu}

\author{Trung Cuong Dang}
\affiliation{
  \institution{University of Central Florida}
  \city{Orlando}
  \state{Florida}
  \country{USA}
}
\email{cuong.dang@ucf.edu}

\author{Corey Pittman}
\affiliation{
  \institution{University of Central Florida}
  \city{Orlando}
  \state{Florida}
  \country{USA}
}
\email{pittmancoreyr@gmail.com}

\author{David Mohaisen}
\affiliation{
  \institution{University of Central Florida}
  \city{Orlando}
  \state{Florida}
  \country{USA}
}
\email{mohaisen@ucf.edu}

\author{Joseph J. LaViola Jr.}
\affiliation{
  \institution{University of Central Florida}
  \city{Orlando}
  \state{Florida}
  \country{USA}
}
\email{jjl@cs.ucf.edu}

\renewcommand{\shortauthors}{Hmaiti et al.}

\begin{abstract}
Privacy measurement instruments (e.g., CFIP, IUIPC, PAQ) predate GDPR by over a decade and measure privacy concerns, distinct from preferences for regulatory protections (e.g., data portability, erasure, automated decision-making rights). This leaves practitioners without tools to assess whether users value the GDPR mechanisms implemented in compliant policies. We developed a GDPR-grounded privacy preference measurement item bank by extracting 669 statements from all 99 GDPR articles, validated by: (1) two-round expert review achieving full consensus on accuracy, (2) semantic clustering into 10 parent themes and 87 subthemes, and (3) consensus review with 50 privacy experts (5 per theme) using a $\geq$4/5 vote retention threshold. The final 527-item bank comprises 9 parent themes and 73 subthemes (18 to 112 items per parent theme, 1 to 29 per subtheme), enabling targeted measurement across granularities while covering GDPR at mean pairwise expert agreement of $\approx$85\%. This work introduces a complementary measurement dimension aligning user preferences with regulatory mechanisms.
\end{abstract}


\maketitle
\pagestyle{plain}
\thispagestyle{plain}

\section{Introduction}
Governments have responded to the rise of the data economy by enacting comprehensive privacy regulations that specify enforceable rights and duties for organizations~\cite{bradford2020brussels,chander2020catalyzing}. The European Union's General Data Protection Regulation (GDPR) is the most influential of these frameworks~\cite{bradford2020brussels}, codifying data subject rights (e.g., access, rectification, erasure, portability, objection/automated decision-making) and processing principles (e.g., lawfulness, fairness, transparency), with extraterritorial scope for entities offering goods/services or monitoring behavior in the EU \cite{bradford2020brussels}. GDPR's influence extends well beyond Europe, inspiring laws like Brazil's LGPD and California's CCPA/CPRA~\cite{chander2020catalyzing}, and catalyzing a global regulatory wave~\cite{bradford2020brussels}.

However, widely used privacy questionnaires were built long before GDPR and primarily measure abstract ``concern'', which is a construct distinct from preferences over specific regulatory protections. Foundational scales like CFIP (1996) and IUIPC (2004) assess dimensions like collection, awareness, and control \cite{smith1996_cfip, malhotra2004_iuipc}. Subsequent analyses show construct ambiguity and limited ability to capture granular preferences \cite{privacy_paradox_meta, colnago2022concern, baruh2017big, wisniewski2016framing}. Hence, studies evaluating user responses to GDPR-era policies and privacy concerns lack instruments designed to assess preferences for the regulation's concrete rights and duties, leaving a gap in conceptual coherence and alignment between measurement tools and regulatory constructs.

Practical repercussions result from this misalignment. Post-GDPR studies document the proliferation of consent notices and privacy disclosures, yet no prevailing instruments were designed to measure whether users value specific consent mechanisms, purposes, or controller obligations \cite{Degeling2019NDSS,Nouwens2020CHI,utz2019informed,matte2020cookie,habib2022okay}. The well-known ``privacy paradox'' persists \cite{privacy_instruments_future_agenda,norberg2007privacy, sundar2020online, wijesekera2018contextualizing, colnago2023there}, but typical measures do not elicit preferences over particular GDPR rights. Representative EU surveys report high awareness of GDPR alongside low exercise of rights~\cite{Eurobarometer2019, strycharz2020data}, and qualitative analyses of data subject access requests (DSAR) workflows describe delays and opacity \cite{bowyer2022humangdpr,habib2020scavenger,dimartino2019leakage, habib2019soups,jamieson2023escaping}. In the absence of regulation-specific measurement tools, it remains unclear whether this is due to implementation barriers or low preference for specific protections (\textsection\ref{sec:legacy_questionnaires}).

Compliance tooling (e.g., commercial checklists and research pipelines) now audits policies against GDPR at scale and standardizes concepts (e.g., DPV) \cite{Harkous2018, Zimmeck2019, Zimmeck2017, Pandit2025DPV, sanchez2019, yan2024quality,Pandit2024semantic}, but we still lack a validated way to measure whether users value those same rights and duties. This asymmetry hampers evidence-based policy and usable-privacy design, and obscures whether organizations' privacy policies and commitments actually match user preferences. Beyond research prototypes, commercial auditors increasingly generate regulator-ready compliance reports for organizations, yet no parallel instruments exist to measure whether end-users value the audited rights and express their preferences towards them when depicted in organizations' privacy policies. 
To our knowledge and after surveying major research venues\footnote{including S\&P, CCS, PETS, CHI, CSCW, SOUPS, USENIX (2018 to 2026) using relevant queries such as \textit{(GDPR OR Data Privacy regulation) AND (scale OR questionnaire OR instrument) AND (preference OR value)}.}, there is no validated, regulation-aligned measurement instrument, which is a distinct construct from the abstract concern measured by existing scales~\cite{colnago2022concern,wisniewski2016framing,baruh2017big}.

To address this gap, we relied on GDPR for item development for three reasons: (1) comprehensiveness and global influence, with uptake in subsequent laws (e.g., LGPD, CCPA/CPRA), (2) maturity of practice since 2018 supporting validation, and (3) article-level specificity enabling systematic extraction of measurement statements. We contribute regulation-aligned, expert-validated ($n=50$) statements to measure GDPR-grounded user privacy preferences. We first \textbf{create} preference statements (initial \(n{=669}\)) covering all 99 GDPR articles, verified for accuracy and clarity through a two-round expert (\(n{= 4}\)) review achieving complete agreement (100\%, $\kappa = 1.0$) on all items. Next, we \textbf{organize} statements into 10 parent themes and 87 subthemes using a data-driven thematic structure, validated for semantic coherence through expert review (\(n{=4}\)). Moreover, we \textbf{establish consensus retention} with a panel of (\(n{= 50}\)) privacy/security experts (5 per parent theme) using an 80\% agreement threshold (\(\geq 4/5\) keep votes), yielding 527 final modular items (mean pairwise expert agreement \(\approx 85\%\)). Lastly, we \textbf{release an open-source searchable database} supporting querying and deployment at multiple granularities (\textsection\ref{sec:usecases} and \textsection\ref{sec:accessSection} for access and supplementary materials), allowing researchers and practitioners to target domains and select response options matching their goals.

By aligning measurement with concrete modern regulatory mechanisms, our items enable studies to test which specific GDPR protections users value, to compare preferences against organizational and data processor policy promises, and to examine whether regulation-specific preference profiles predict GDPR-related behaviors, representing a hypothesis that legacy concern scales were not designed to test. We argue that our work contributes to reducing the methodological gap between modern compliance practice and user-centric measurement. Our item bank is designed for modular deployment and adaptation (\textsection\ref{sec:usecases}) and can be used across different contexts, e.g., lab studies, survey, field contexts to inform user privacy, policy, compliance, privacy behavior, product decisions, etc.

\vspace{-2mm}
\section{Background and Related Work}
Here, we review existing scales and their temporal misalignment (\textsection\ref{sec:legacy_questionnaires}), GDPR's distinction between user rights and controller principles (\textsection\ref{sec:gdpanalysis_rw}), automated policy analysis showing the measurement gap (\textsection\ref{sec:taxandautomate}), and expert review methods guiding our protocol (\textsection\ref{sec:delphi_rw}).

\vspace{-2mm}
\subsection{Data Privacy Preferences Measurement}
\label{sec:legacy_questionnaires}
\begin{table}[t]
\centering
\caption{Privacy Instruments: Scope Comparison Across Concern and Preference Constructs}
\label{tab:instrument_comparison}\vspace{-2mm}
\scriptsize
\setlength{\tabcolsep}{1.2pt}
\begin{tabular}{@{}lcccccccc@{}}
\toprule
\textbf{Instrument} & \textbf{Year} & \textbf{Items} & \textbf{Construct} &
\rotatebox{60}{\makecell{\textbf{GDPR}\\\textbf{Rights}}} &
\rotatebox{60}{\makecell{\textbf{GDPR}\\\textbf{Principles}}} &
\rotatebox{60}{\makecell{\textbf{Article}\\\textbf{Mapped}}} &
\rotatebox{60}{\textbf{Modular}} &
\rotatebox{60}{\makecell{\textbf{Open}\\\textbf{Access}}} \\
\midrule
CFIP~\cite{smith1996_cfip} & 1996 & 15 & Concern & \xmark & \pmark & \xmark & \xmark & \cmark \\
IUIPC~\cite{malhotra2004_iuipc} & 2004 & 10 & Concern & \xmark & \pmark & \xmark & \xmark & \cmark \\
IPC~\cite{dinev2006_internet_privacy} & 2006 & 4\textsuperscript{†} & Concern & \xmark & \pmark & \xmark & \xmark & \cmark \\
OPC~\cite{buchanan2007_opc} & 2007 & 16 & Concern & \xmark & \pmark & \xmark & \xmark & \cmark \\
MUIPC~\cite{xu2012measuring} & 2012 & 9 & Concern & \xmark & \pmark & \xmark & \xmark & \cmark \\
\midrule
\rowcolor{gray!15}
\textbf{GPPI (ours)} & \textbf{2026} & \textbf{527} & \textbf{Preference} & \textbf{\cmark} & \textbf{\cmark} & \textbf{\cmark} & \textbf{\cmark} & \textbf{\cmark} \\
\bottomrule
\end{tabular}
\begin{minipage}{\columnwidth}
\scriptsize
\vspace{2pt}
Legacy scales measure privacy concern, not regulatory preference; the comparison highlights complementary scope rather than superiority. 
\textbf{Construct:} Concern = general privacy worry; Preference = valuation of specific protections.  
\textbf{GDPR Coverage:} Covers GDPR rights, principles, obligations, transfers, and enforcement mechanisms. \textbf{\pmark}~= related pre-GDPR concepts (e.g., collection, control).  
\textbf{Article Mapped:} Items linked to specific GDPR provisions.  
\textbf{Modular:} Supports domain-specific deployment and targeted queries (9 themes, 73 subthemes).  
\textbf{Open Access:} Item bank and code publicly available.  
\textbf{Legend:} \textbf{\cmark~= Yes; \pmark~= Partial/related; \xmark~= No.}  
\textsuperscript{†}IPC item count varies across studies (4 items in the original study; 6+ in later work).
\end{minipage}\vspace{-4mm}
\end{table}

Early work on privacy preferences centered on attitudinal indices such as Westin's Privacy Segmentation Index, which classifies individuals as ``fundamentalists,'' ``pragmatists,'' or ``unconcerned'' \cite{westin_privacy_index, westin1979dimensions, kumaraguru2005privacy}. Subsequent studies questioned how well such coarse categories predict real-world disclosure choices and configuration behaviors, finding weak attitude-behavior alignment in many settings~\cite{privacy_paradox_meta, westin_critique_behavior, dienlin2015privacy}. To capture finer-grained concerns, in 1996, Smith et al. introduced the ``Concern for Information Privacy'' (CFIP) scale~\cite{smith1996_cfip} with four key dimensions: collection, unauthorized secondary use, improper access, and data errors. In 2004, Malhotra et al.  proposed IUIPC, a second-order instrument emphasizing control, awareness, and collection, for online contexts \cite{malhotra2004_iuipc}. Dinev \& Hart's Internet privacy concern measures (2006) foreground concerns about information abuse and finding~\cite{dinev2006_internet_privacy}, while Buchanan et al. developed the OPC scale in 2006 with components for general caution, technical protection behaviors, and overall privacy-related attitude \cite{buchanan2007_opc}. Xu et al.'s MUIPC (2012) extended privacy concern measurement to mobile contexts~\cite{xu2012measuring}.

While widely used, follow-on assessments noted construct and validity issues in some sub-scales and contexts, suggesting room for refinement and revalidation \cite{iuipc_reliability_reexam, cfip_update_validity}. IUIPC and OPC capture only partially overlapping constructs \cite{buchanan2007_opc}, and recent work shows users often misinterpret statements intended for measuring privacy \textit{concerns} as privacy \textit{preferences}, with no statement uniquely capturing any single construct~\cite{colnago2022concern}.
With the multifaceted nature of ``privacy concern'', existing privacy scales developed prior to modern regulatory frameworks suffer from construct ambiguity and emphasize latent dimensions \cite{privacy_measurement_review_modern, privacy_paradox_meta}, and were not designed to assess preferences for specific regulatory protections that now substantively shape users' real-world privacy experiences and choices.

The regulatory landscape has shifted substantially since foundational privacy scales were developed. For example, GDPR (effective 2018) codifies concrete subject rights and processing principles (e.g., erasure, access, portability, purpose limitation, lawfulness) \cite{gdpr_text_2016, Linden2020GDPRLandscape, solove2016brief}, while the CCPA/CPRA (effective 2020/2023) introduces opt-out of sale/sharing and additional consumer rights \cite{ccpa2018,cpra2020}. Instruments created in the 1990s and 2000s do not directly assess preferences for these specific legal protections. This creates a construct misalignment: studies of users’ aptitudes, preferences, and reactions to GDPR/CCPA-compliant policies often rely on scales measuring broad privacy ``concern'' or attitudes, which capture a related but distinct construct from preferences over the rights and policies these regulations implement and enforce, weakening conceptual coherence, predictive validity, and generalization~\cite{colnago2022concern,privacy_instruments_future_agenda,privacy_paradox_meta}.

Recent compliance studies indicate that GDPR-compliant policies now prominently feature regulatory concepts aligned with GDPR-era rights and duties \cite{Linden2020GDPRLandscape}, yet existing questionnaires and tools do not target the measurement of constructs now central to users' legal protections and policy experiences. Because no existing tools measure user preferences for specific regulatory provisions, practical consequences arise: companies design regulation-compliant policies addressing regulatory requirements (e.g., GDPR, the focus of our research), but user aptitudes, priorities and preferences towards these cannot be measured by existing tools. This gap between regulations, modern privacy policies, and preference measurement tools motivates creating regulation-aligned measurement items connecting user data privacy preferences to concrete regulatory constructs and observable decisions \cite{privacy_instruments_future_agenda}. Figure~\ref{fig:positioning} summarizes this construct gap and positions GPPI as a GDPR-grounded complement to legacy privacy-concern scales.

\definecolor{blueborder}{RGB}{110, 158, 205}
\definecolor{bluefill}{RGB}{240, 247, 253}
\definecolor{greenborder}{RGB}{88, 158, 108}
\definecolor{greenfill}{RGB}{238, 250, 240}
\definecolor{orangeborder}{RGB}{205, 148, 88}
\definecolor{orangefill}{RGB}{253, 245, 235}
\definecolor{textgray}{RGB}{55, 55, 55}

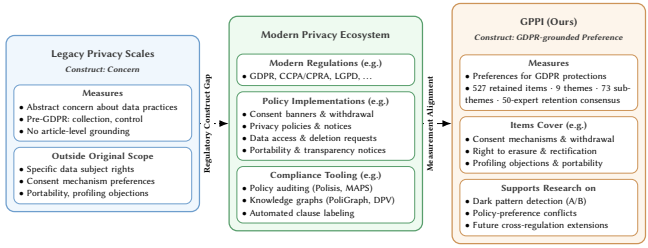
\begin{figure}[t]
\centering
\resizebox{\columnwidth}{!}{%
\begin{tikzpicture}[
    font=\sffamily,
    >=Latex,
    node distance=0.28cm and 0.9cm,
    title/.style={font=\bfseries\large, text=textgray, align=center},
    gap_arrow/.style={-{Latex[length=3mm]}, thick, dashed, draw=black},
    bridge_arrow/.style={-{Latex[length=3mm]}, thick, draw=black}
]

\node[draw=blueborder, fill=bluefill, rounded corners=8pt,
      very thick, inner sep=7pt] (leftpanel)
{
  \begin{tikzpicture}[node distance=0.18cm]
    \node[title] (ltitle) {
      Legacy Privacy Scales\\[0.5mm]
      {\small\itshape\color{textgray} Construct: Concern}
    };
    \node[draw=blueborder, fill=white, rounded corners=6pt, line width=0.8pt,
          text width=5.6cm, align=left, inner sep=5pt, below=of ltitle] (l1) {
      \makebox[\linewidth][c]{\bfseries\color{textgray} Measures}\\[1.5pt]
      $\bullet$ Abstract concern about data practices\\[0.5pt]
      $\bullet$ Pre-GDPR: collection, control\\[0.5pt]
      $\bullet$ No article-level grounding
    };
    \node[draw=blueborder, fill=white, rounded corners=6pt, line width=0.8pt,
          text width=5.6cm, align=left, inner sep=5pt, below=of l1] (l2) {
      \makebox[\linewidth][c]{\bfseries\color{textgray} Outside Original Scope}\\[1.5pt]
      $\bullet$ Specific data subject rights\\[0.5pt]
      $\bullet$ Consent mechanism preferences\\[0.5pt]
      $\bullet$ Portability, profiling objections
    };
  \end{tikzpicture}
};

\node[draw=greenborder, fill=greenfill, rounded corners=8pt,
      very thick, inner sep=7pt, right=0.95cm of leftpanel] (centerpanel)
{
  \begin{tikzpicture}[node distance=0.18cm]
    \node[title] (ctitle) {Modern Privacy Ecosystem};
    \node[draw=greenborder, fill=white, rounded corners=6pt, line width=0.8pt,
          text width=5.6cm, align=left, inner sep=5pt, below=of ctitle] (c1) {
      \makebox[\linewidth][c]{\bfseries\color{textgray} Modern Regulations (e.g.)}\\[1.5pt]
      $\bullet$ GDPR, CCPA/CPRA, LGPD, \ldots
    };
    \node[draw=greenborder, fill=white, rounded corners=6pt, line width=0.8pt,
          text width=5.6cm, align=left, inner sep=5pt, below=of c1] (c2) {
      \makebox[\linewidth][c]{\bfseries\color{textgray} Policy Implementations (e.g.)}\\[1.5pt]
      $\bullet$ Consent banners \& withdrawal\\[0.5pt]
      $\bullet$ Privacy policies \& notices\\[0.5pt]
      $\bullet$ Data access \& deletion requests\\[0.5pt]
      $\bullet$ Portability \& transparency notices
    };
    \node[draw=greenborder, fill=white, rounded corners=6pt, line width=0.8pt,
          text width=5.6cm, align=left, inner sep=5pt, below=of c2] (c3) {
      \makebox[\linewidth][c]{\bfseries\color{textgray} Compliance Tooling (e.g.)}\\[1.5pt]
      $\bullet$ Policy auditing (Polisis, MAPS)\\[0.5pt]
      $\bullet$ Knowledge graphs (PoliGraph, DPV)\\[0.5pt]
      $\bullet$ Automated clause labeling
    };
  \end{tikzpicture}
};

\node[draw=orangeborder, fill=orangefill, rounded corners=8pt,
      very thick, inner sep=7pt, right=0.95cm of centerpanel] (rightpanel)
{
  \begin{tikzpicture}[node distance=0.18cm]
    \node[title] (rtitle) {
      GPPI (Ours)\\[0.5mm]
      {\small\itshape\color{textgray} Construct: GDPR-grounded Preference}
    };
    \node[draw=orangeborder, fill=white, rounded corners=6pt, line width=0.8pt,
          text width=5.6cm, align=left, inner sep=5pt, below=of rtitle] (r1) {
      \makebox[\linewidth][c]{\bfseries\color{textgray} Measures}\\[1.5pt]
      $\bullet$ Preferences for GDPR protections\\[0.5pt]
      $\bullet$ 527 retained items $\cdot$ 9 themes $\cdot$ 73 subthemes $\cdot$ 50-expert retention consensus
    };
    \node[draw=orangeborder, fill=white, rounded corners=6pt, line width=0.8pt,
          text width=5.6cm, align=left, inner sep=5pt, below=of r1] (r2) {
      \makebox[\linewidth][c]{\bfseries\color{textgray} Items Cover (e.g.)}\\[1.5pt]
      $\bullet$ Consent mechanisms \& withdrawal\\[0.5pt]
      $\bullet$ Right to erasure \& rectification\\[0.5pt]
      $\bullet$ Profiling objections \& portability
    };
    \node[draw=orangeborder, fill=white, rounded corners=6pt, line width=0.8pt,
          text width=5.6cm, align=left, inner sep=5pt, below=of r2] (r3) {
      \makebox[\linewidth][c]{\bfseries\color{textgray} Supports Research on}\\[1.5pt]
      $\bullet$ Dark pattern detection (A/B)\\[0.5pt]
      $\bullet$ Policy-preference conflicts\\[0.5pt]
      $\bullet$ Future cross-regulation extensions
    };
  \end{tikzpicture}
};

\draw[gap_arrow] (leftpanel.east) --
    node[midway, above, font=\bfseries\small, rotate=90,
         fill=white, draw=none, inner sep=2pt]
    {Regulatory Construct Gap} (centerpanel.west);

\draw[bridge_arrow] (centerpanel.east) --
    node[midway, above, font=\bfseries\small, rotate=90,
         fill=white, draw=none, inner sep=2pt]
    {Measurement Alignment} (rightpanel.west);

\end{tikzpicture}%
}\vspace{-2mm}\caption{GPPI complements legacy privacy-concern scales by measuring preferences toward concrete GDPR protections reflected in modern privacy policies.}
\label{fig:positioning}\vspace{-7mm}
\end{figure}

\vspace{-2mm}
\subsection{GDPR and Legal Framework Analysis}
\label{sec:gdpanalysis_rw}
The introduction of GDPR changed the privacy landscape, yet research examining its impact faces a challenge: \textit{how do we assess user preferences for the specific regulatory protections and rights that GDPR introduced and that organizations and data processors implement in their policies?} We identify a consistent gap in measurement capabilities that limits our understanding of GDPR's effectiveness.

Prior implementation studies indicate unmeasured preference gaps. Since GDPR took effect, empirical studies examined how organizations operationalize its requirements. Large-scale measurements report more prevalent privacy notices and consent banners post-enforcement, yet third-party tracking practices changed only modestly, suggesting that many deployments satisfy the requirements of the law more than its spirit~\cite{Degeling2019NDSS}. 
Subsequent work shows that consent interfaces often use nudging/dark patterns, and small UI changes materially shift acceptance (e.g., removing ``Reject all'' on the first page increases acceptance by $\approx$22{}-{}23\%; adding granular controls reduces it by $\approx$8-20\%), undermining the ideal of ``freely given'' and informed consent~\cite{Nouwens2020CHI}. These studies document implementation patterns but do not assess whether users actually prefer the specific consent mechanisms, data processing purposes, or controller obligations that GDPR mandates, limiting our understanding of user-regulation alignment and user-regulation-policy alignment.

We find privacy paradox research lacks regulatory specificity, and decades of research document gaps between stated privacy attitudes and observed behavior~\cite{norberg2007privacy, privacy_instruments_future_agenda}. Post-GDPR studies continue to uncover design manipulations (e.g., nudging/dark patterns) and indicate that individuals express concern yet often disclose due to different motives, such as convenience or social reasons, indicating that legal rights alone do not close the intention-behavior gap \cite{utz2019informed, machuletz2019multiple, nguyen2021share, nguyen2022}. However, these studies measure generic privacy concern, a construct that does not capture preferences for specific GDPR rights (e.g., data portability, erasure, etc.) or processing principles (e.g., lawfulness, fairness, etc.). Without measuring preferences for these protections, we cannot determine whether the paradox persists because users do not value the GDPR provisions or because implementation barriers prevent acting on preferences.

Existing studies also cannot distinguish whether low GDPR exercise rates stem from implementation barriers or from genuinely low preference for these rights, as no widely adopted prior instruments are aligned and validated for modern regulations, operationalizing preferences over specific GDPR rights. Representative EU surveys show high awareness of GDPR and its rights, but far fewer people actually exercise them: the 2019 Eurobarometer found 67\% had heard of GDPR and 73\% knew at least one right, yet only 24\% had ever objected to direct marketing, 18\% had requested access, and 16\% had requested correction~\cite{EU2019_EB487a}. Strycharz et al. reported similar patterns of awareness without corresponding action~\cite{strycharz2020data}. This gap aligns with findings that widely used privacy scales often capture generic privacy attitudes and exhibit construct ambiguity rather than directly eliciting preferences over specific rights~\cite{colnago2022concern}. Qualitative studies further indicate implementation barriers (delayed, incomplete, or opaque DSAR responses) that leave users feeling disempowered and help explain low exercise rates \cite{bowyer2022humangdpr, pohn2025qualitative}.

Across the implementation, paradox, and privacy-rights exercise literature, researchers rely on ad hoc measures that lack standardization and alignment with modern data-protection frameworks~\cite{colnago2022concern, Linden2020GDPRLandscape}. This gap creates an attribution challenge: observed patterns may reflect genuine user preferences, implementation issues (e.g., consent/UI design, DSAR barriers), or measurement artifacts. As a result, the field cannot determine whether regulations like the GDPR empower users or fall short due to barriers, an important distinction for evidence-based policy reform. The implications extend beyond compliance to organizational strategy: when companies satisfy GDPR requirements, regulation-aligned preference instruments, building on validated items can reveal mismatches between user values and implemented policies, enabling targeted improvements to both regulatory and organizational practices.

\vspace{-2mm}
\subsection{Policy Taxonomies and Automated Analysis}
\label{sec:taxandautomate}
Prior work systematized privacy policies with taxonomies, corpora, and automated analysis. A foundational corpus, \textit{OPP-115}, introduced fine-grained annotations across ten data-practice categories and enabled learning-based segment classification \cite{Wilson2016Corpus}. Building on such corpora, \textit{Polisis} trained privacy-tailored models over large policy collections to classify practices, answer policy questions, and generate standardized ``nutrition labels'' \cite{Harkous2018}. At scale, compliance-oriented pipelines link policy text to observed behavior (e.g., \textit{MAPS} for Android apps), indicating gaps between stated practices and implementations \cite{Zimmeck2019}. Longitudinal measurements document GDPR-era shifts in policy content and readability (e.g., increased disclosures of rights and international transfers) \cite{Linden2020GDPRLandscape}. Human-Computer Interaction (HCI) research complements these efforts by indicating that standardized, label-like presentations improve users' ability to locate data practices, highlighting the value of simplified policy representations for end users \cite{Kelley2010}.

Recent work modernizes analysis. GDPR-aligned ontologies such as the Data Privacy Vocabulary (DPV) standardize concepts (purposes, legal bases, rights) for machine-readable analysis \cite{Pandit2025DPV}. Knowledge-graph extraction (\textit{PoliGraph}) represents cross-sentence data-collection statements as relations among entities, data types, and purposes \cite{Cui2023PoliGraph}. LLM-based pipelines scale clause-level labeling across jurisdictions and large site cohorts \cite{Xie2025LLMPolicies}. Domain-focused quality auditing (e.g., \textit{QuPer} for virtual assistants) operationalizes policy timeliness, availability, completeness, and readability, finding that $\approx$1\% of policies fully cover GDPR-required content \cite{yan2024quality}. Tooling also standardizes policy collection and detection across languages, improving dataset hygiene for analyses \cite{Hosseini2021Detector}.

User-facing policy tools increasingly focus on improving comprehension rather than supporting text mining. Early approaches like interactive policy presentations, increase perceived transparency and engagement without inflating trust or perceived control \cite{Reinhardt2021VIPP}. Building on the idea of supporting users at decision point, contextual injection of policy snippets into web pages (\textit{PrivacyInjector}) fosters more reflective privacy behaviors in lab and field settings \cite{Windl2022PrivacyInjector}. Complementing these interface-level interventions, browser-based summarizers (e.g., \textit{PrivacyCheck~v2}) map policies to structured Q\&A about GDPR rights to improve user understanding \cite{Zaeem2020PrivacyCheck}. Community-driven efforts such as ``Terms of Service; Didn't Read’’ (ToS;DR, see: https://tosdr.org) provide crowdsourced ratings and plain-language summaries that help users quickly assess service practices. More automated approaches, such as TLDR, use machine-learning–based annotation to categorize policy content and highlight key segments, significantly reducing reading time while improving comprehension~\cite{mohaisen2021tldr}. Recently, LLM-powered summarization interfaces (\textit{Privacify}) extended these capabilities by generating plain-language disclosures with explainability features \cite{Woodring2024Privacify}.

While automated policy analysis can extract and annotate regulation relevant concepts from policies (e.g., data rights, processing purposes, legal bases), no validated instruments exist to measure user preferences for these same regulatory themes. This further solidifies the asymmetry: we can automatically detect what policies promise about GDPR compliance, but cannot assess whether a user actually values these protections, understands their implications, or approve exceptions to these protections in a provider’s policy. This motivates the creation of regulation-grounded preference measurement tools that align with privacy regulation themes.

\vspace{-2mm}
\subsection{Expert Review and GPPI Development}
\label{sec:delphi_rw}
To validate new measurement instruments, security/privacy and HCI research combines structured review for content validity with expert consensus methods. Within security, privacy, and HCI domains, expert panels have been used to prioritize risks and to produce artifacts and frameworks (e.g., ranked research agendas, patient-facing explanations, organizational readiness components) \cite{haynes2021delphi,schmit2020delphi,OBrien2021,Sherman2017CATSDelphi}. For item-level vetting, single- or two-phase expert ratings (e.g., relevance, clarity, coverage) are quantified via the Content Validity Index/Ratio (CVI/CVR), with items revised or removed according to apriori thresholds~\cite{Lynn1986,PolitBeck2006}. Guided by this literature, we use a two-stage process: (1) expert review to screen and edit items for GDPR alignment and clarity, and (2) an expert review round with controlled feedback to finalize inclusion/exclusion and wording and document consensus.

\begin{figure*}
    \centering
    \includegraphics[width=.9\linewidth]{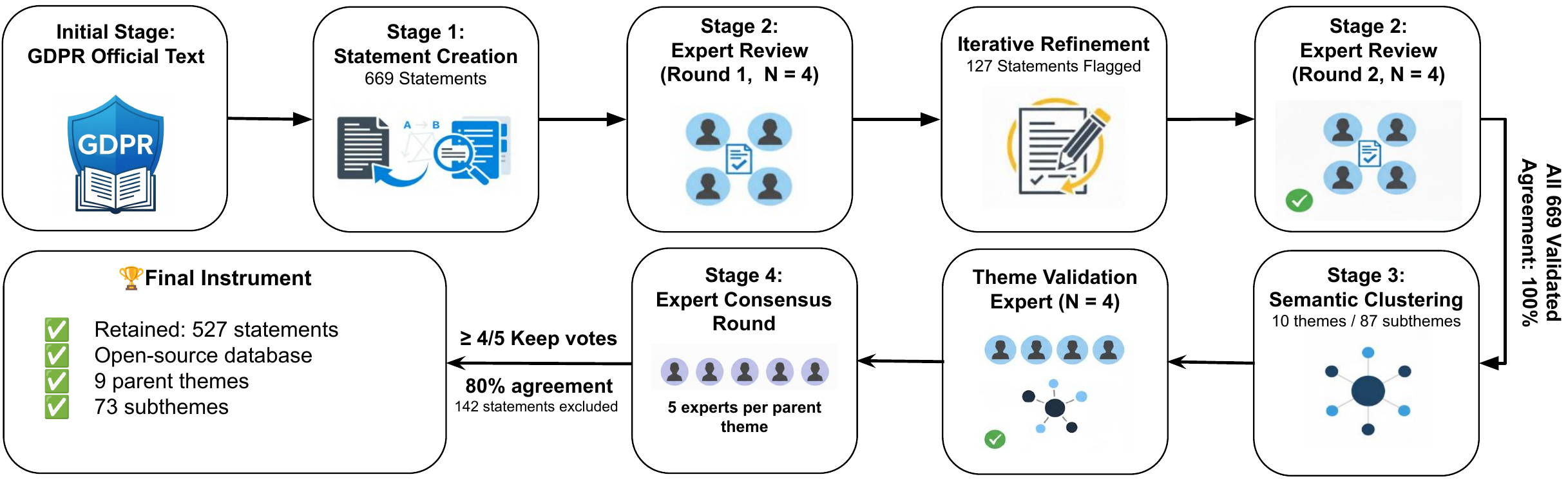}\vspace{-2mm}
    \caption{Pipeline for building GPPI. Starting from GDPR text, we created 669 statements (Stage 1). Two expert-review rounds ($n = 4$) achieved 100\% agreement after revising flagged statements (Stage 2). Statements were semantically clustered into 10 parent themes and 87 subthemes and validated by experts (Stage 3). A consensus process with 50 experts retained items receiving $\geq 4/5$ (80\%) ``keep'' votes (Stage 4), yielding 527 final statements (142 removed).}
    \label{fig:methodology}\vspace{-2mm}
\end{figure*}

\vspace{-2mm}
\section{GPPI Development and Validation}
\label{sec:methodology}
Our methodology ensures that retained statements are: (1) regulatory-accurate (\textsection\ref{sec:stage1}--\textsection\ref{sec:stage2}), (2) organized into semantically coherent themes (\textsection\ref{sec:stage3}), and (3) validated by experts as clear, non-redundant, and regulation-grounded (\textsection\ref{sec:stage4}). We extracted 669 statements from all 99 GDPR articles (\textsection\ref{sec:stage1}), achieved 100\% expert agreement on regulatory accuracy through two-round validation by four experts (\textsection\ref{sec:stage2}), organized statements into 10 parent themes and 87 subthemes through semantic clustering (\textsection\ref{sec:stage3}), and retained 527 statements meeting $\geq$80\% consensus among 50 experts (\textsection\ref{sec:stage4}). All procedures were approved by our Institutional Review Board (IRB). Appendix~\ref{app:sample} shows two statements across all four stages, including extraction, expert revision, thematic assignment, and retention/removal.

\vspace{-2mm}
\subsection{Stage 1: Statement Creation}
\label{sec:stage1}
We extracted and transformed official GDPR text~\cite{gdpr_text_2016} to user-comprehensible preference statements to ensure comprehensive coverage while maintaining legal accuracy and representation. This process involved structured parsing, cross-reference integration, user-facing transformation, and refinement.

We parsed the complete GDPR text (Regulation EU 2016/679) following its hierarchical structure: Chapter $\rightarrow$ Section $\rightarrow$ Article $\rightarrow$ Segment. Each segment represents the smallest regulatory unit within an article, with prescriptive requirements, definitions, or procedural specifications. We created items exclusively from all GDPR's enacting terms (Articles), omitting the preamble and EUR-Lex metadata. Recitals were consulted only to clarify meaning where an Article was ambiguous, consistent with EU guidance and case law that treats recitals as interpretive aids rather than bindings. 

We applied rules to create the remaining statements, especially for clauses with bullet points. \textbf{Disjunctive segments (``OR'')} were treated as discrete alternatives, split into separate statements with repeating governing preposition, subject, or condition, so each item is contextualized and can stand alone. \textbf{Conjunctive segments (``AND'')} were concatenated into a single statement to preserve the regulation's meaning and ensure that listed conditions apply simultaneously. \textbf{When a segment contained cross-references} to other GDPR provisions or external sources, we resolved the reference by extracting the relevant context and integrating it inline. This produced self-contained statements that respondents can interpret without navigating cross-references, while maintaining fidelity to the scope, context, obligations, and exceptions.

Regulatory language is written for legal practitioners and data controllers, not users providing privacy preferences~\cite{waldman2018privacy, becher2021readability,jensen2004evaluation}. To create accessible preference statements, we followed a transformation process prioritizing clarity, accessibility, fidelity, and reduction of cognitive load as recommended by prior work~\cite{habib2022usabilityofprivacychoice, kelley2009label, seizov2019transparent}. We transformed each regulatory segment into a first-person preference statement following these principles:
\label{sec:formatting}

\begin{itemize}\vspace{-1.5mm}
    \item \textbf{Consistent framing:} all statements began with ``I would like my personal data to be processed only if\ldots'' to ensure consistent expert review as seen by end users, aligning expert judgments with deployment. Responses focus on privacy trade-offs (agree/disagree/ask-me-later), treating users as decision-makers rather than notice recipients~\cite{schaub2015designspace, 
habib2022usabilityofprivacychoice}. The prefix and response are adaptable to specific use cases (\textsection\ref{sec:usecases}).
    \item \textbf{Plain language:} legal terms were translated to accessible terms while preserving regulatory meaning~\cite{waldman2018privacy, becher2021readability}. To preserve legal precision, we kept some terms and added clarification.    
    \item \textbf{Completeness:} Statements captured the scope of requirements, including conditional clauses, exceptions, and safeguards, to avoid misrepresenting rights or controller obligations.
    \item \textbf{Neutrality:} Statements used balanced formulations that present privacy trade-offs without implicit value judgments~\cite{privacy_instruments_future_agenda}, to avoid leading or biased phrasing that might influence user responses.
\end{itemize}

The creation and translation process was conducted by three expert doctoral students, where each followed GDPR linearly, producing independent statement formulations following the above guidelines. This resulted in an equal number (669) of statement formulations per researcher, and differences between formulations were reconciled during synchronous meetings to produce a unified statement bank. Disagreements were resolved through discussion of regulatory intent, consultation of GDPR for clarification, and consideration of how data subjects would interpret the requirement in practice. The unified statements were organized in a structured spreadsheet with columns: \textit{Chapter}, \textit{Section}, \textit{Article}, \textit{Original Segment}, \textit{References (if any)}, and \textit{GPPI Statement}. The sheet was set-up so that each statement retained explicit traceability to its source regulatory provision, with external references embedded as links and ones internal to GDPR as text. The resulting spreadsheet of 669 self-contained preference statements spanning all 99 GDPR articles formed the basis for expert accuracy validation (Stage 2, \textsection\ref{sec:prevalidation}).

\vspace{-2mm}
\subsection{Stage 2: Expert Accuracy Validation}
\label{sec:prevalidation}
\label{sec:stage2}
We conducted a two-round validation to ensure all candidate statements accurately represented GDPR requirements, correctly incorporated references (if present in original segment) and were clearly worded. This validation stage served two purposes: eliminate fundamental accuracy issues that would confound consensus decisions, and establish baseline inter-rater reliability agreement (IRR) to confirm experts could consistently evaluate GDPR-aligned statements and agree on their appropriateness and formulation. We recruited four independent domain experts to validate all 669 statements for regulatory accuracy and user comprehensibility prior to large-scale consensus review. We recruited experts via direct outreach to researchers identified from GDPR, privacy measurement, usable privacy, and privacy law literature (2018--2025). Of 60 contacted, 4 agreed to participate in this 12-week hands-on validation. 

All experts met these criteria: (1) terminal degree in computer science or related field with focus on security/privacy, (2) $\geq$5 years post-PhD experience, (3) $\geq$5 first-authored publications in A* privacy/security/HCI venues, (4) demonstrated GDPR expertise through publications, teaching, or professional practice, and (5) fluent English proficiency. The final panel comprised two security/privacy researchers and two HCI researchers specializing in usable privacy (see Appendix~\ref{app:experts}). All experts voluntarily participated and opted to have their names/affiliations acknowledged post-acceptance.

\subsubsection{Two-Round Protocol with Four Domain Experts}
We used a structured evaluation protocol to ensure consistency across reviewers and facilitate systematic revision. Each expert received the complete statement bank (669 items) in a standardized spreadsheet containing: \textit{statement ID} (unique identifier for tracking), \textit{statement text} (user-facing preference statement), \textit{original regulatory segment} (exact GDPR language for comparison), \textit{GDPR sources} (Chapter, Section, Article, Page), and \textit{references} (any cross- or external-references integrated from cited GDPR articles or external regulations). Experts completed reviews independently without consultation to ensure unbiased assessments. We explicitly instructed them to prioritize regulatory accuracy and clarity to users over brevity, considering that some GDPR provisions may need more details. Per statement, experts provided the following:
\begin{itemize}
    \item \textbf{Accuracy assessment:} binary decision (Agree/Disagree) on whether the statement correctly and completely represents the original GDPR segment.
    \item \textbf{Justification (required for Disagree):} open-ended specific explanation of inaccuracy.
    \item \textbf{Suggested revision (required for Disagree):} Concrete alternative wording that addresses the identified issue while maintaining user comprehensibility
    \item \textbf{Optional comments:} any additional observations.
\end{itemize}

This evaluation cycle took 3 weeks. Across 669 statements, we observed IRR agreement: \textbf{unanimous agreement (4/4 Agree):} 542 statements (81.0\%), \textbf{majority agreement (3/4 Agree):} 89 statements (13.3\%), \textbf{split or minority agreement ($\leq$2/4 Agree):} 38 statements (5.7\%). Overall percentage agreement was 88.5\% (2,373 Agree ratings out of 2,676 total evaluations), indicating strong consensus despite the complexity of GDPR requirements. The 127 flagged statements concentrated in provisions with conditional exceptions, dense cross-references, or scope ambiguity. For disagreements, we coded all 127 flagged statements (at least one disagree) according to expert justifications. We identified three issue types: \textbf{incomplete representation (76 statements, 60\%)}, where statements omitted conditions, exceptions, or safeguards present in the source article. \textbf{Unclear transformation (36 statements, 28\%)}, where legal terminology was translated ambiguously or could confuse users. \textbf{Incorrect scope interpretation (15 statements, 12\%)}, where statements described  requirements too broadly or narrowly relative to their actual GDPR scope. We then conducted discussions involving all authors to revise statements following these criteria: for \textbf{single expert disagreements}, the lead author reviewed the concern against the original GDPR provision in consultation with all co-authors; any concern reflecting substantive regulatory inaccuracy, rather than formatting or phrasing preference, triggered revision, ensuring no genuine accuracy issue was dismissed by majority override. For \textbf{multiple expert disagreements}, we consulted the GDPR text and our statement to synthesize expert suggestions into a unified revision addressing all identified concerns.

We revised all 127 flagged statements by re-examining GDPR source text, synthesizing expert recommendations, and addressing all identified concerns while maintaining accessibility. Expert revision suggestions never conflicted, meaning multiple experts flagging the same statement proposed consistent fixes. Revisions were done through synchronous meetings with all research team members, and prioritized completeness and accuracy over brevity. The revised statements followed the statement transformation and format described in \textsection\ref{sec:formatting}.
We returned the complete revised statement bank (669 items, 127 revised) to the four experts independently with instructions to evaluate all revised statements (clearly marked rows with a background color), confirm whether revisions adequately addressed Round 1 concerns, flag any remaining accuracy issues using the same Agree/Disagree protocol with justifications for disagreement, and optionally review statements agreed on from Round 1 if new concerns emerged. 

Round 2 confirmed successful convergence: all four experts rated all 669 statements as ``Agree'', achieving 100\% agreement ($\kappa = 1.0$). No expert identified new concerns in previously agreed statements. This validation established the statement bank's regulatory accuracy and user comprehensibility, enabling thematic organization (\textsection\ref{sec:clustering}) without confounding quality issues. 

\vspace{-2mm}
\subsection{Stage 3: Clustering \& Themes Formation}
\label{sec:clustering}
\label{sec:stage3}
GDPR's structure is designed for legal interpretation and compliance, not user-preference measurement. Articles are grouped by regulatory function rather than themes reflecting a data subject's perspective. Key concepts cut across the text, e.g., \textit{consent} appears in Art.~4(11) (definition), 6(1)(a) (lawful basis), 7 (conditions for consent), 8 (conditions applicable to child's consent in relation to information society services), 9(2)(a) (processing of special categories of personal data), and 13-14 (transparency). Also, article numbering does not encode conceptual proximity, e.g., Art.~6 (lawfulness of processing) and Art.~32 (security of processing) impose controller obligations yet are apart. This poses challenges: statements should be grouped by \emph{semantic similarity} (what preference is expressed) and \emph{regulatory coherence} (which obligation it addresses).

Pure text-based clustering ignores legal context, whereas organizing strictly by Articles reproduces a legalistic layout and misses cross-cutting themes. Therefore, we applied a data-driven semantic clustering to reorganize the 669 validated statements into coherent themes to aid respondent understanding. This also allowed cross-cutting regulatory concepts to emerge naturally while maintaining complete coverage and traceability to source provisions.

\subsubsection{Feature Engineering and Clustering Pipeline}
To capture semantic similarity and regulatory context, we constructed feature representations combining three information sources: \textit{statement semantic content}, \textit{conceptual keywords}, and \textit{GDPR regulatory context}.

\textbf{Statement Semantic Embeddings:} we encoded each statement using \textit{all-MiniLM-L6-v2} (384-d) from Sentence-Transformers~\cite{allminilm6v2, reimers2019sentencebertsentenceembeddingsusing, wang2020minilm, muennighoff2022mteb}, a lightweight model for semantic similarity and clustering at our scale. This captures meaning, enabling identification of related statements despite phrasing differences.

\textbf{Keyword Semantic Embeddings:} To isolate core privacy concepts from syntactic variation, three research team members independently extracted five keywords per statement capturing: (1) primary GDPR principle/right (e.g., consent, erasure), (2) regulatory actor (e.g., controller, supervisory authority), (3) operational context (e.g., automated decision-making, cross-border), and (4) relevant additional information. Disagreements were resolved through structured discussion, producing a consensus keyword ($n = 5$) list per statement. Each statement's five keywords were encoded using the same sentence-transformer and mean-pooled into 384-dimensional representations, ensuring statements sharing keywords and core concepts cluster together despite structural differences.

\textbf{Regulatory Context Embeddings:} Each statement was accompanied by metadata from its source GDPR provision (Chapter, Section, Article). These concatenated regulatory descriptors (index/name) were encoded using the same sentence-transformer (384-d) context embeddings, preserving GDPR's hierarchical structure while allowing semantically coherent cross-cutting themes to emerge when content similarity outweighs regulatory proximity.

\textbf{Feature Concatenation:} The three embedding types were concatenated into unified 1,152-dimensional feature vectors. Features were normalized using a z-score to ensure equal contribution from each modality during clustering. Ablation analysis (See Appendix) indicated that all features concatenated outperformed any single type representation in cluster coherence metrics (Silhouette Score).

\textbf{Projection \& Clustering:} The 1,152-dimensional feature space was projected to 2D using UMAP~\cite{mcinnes2018umap} (configured with cosine similarity, 10 neighbors, 0.1 minimum distance, random seed 42). UMAP preserves both local and global structure~\cite{mcinnes2018umap}, enabling interpretable clustering that captures GDPR's hierarchical organization while allowing cross-cutting themes to emerge. We evaluated three clustering approaches on the 2D UMAP space: k-means~\cite{lloydkmeans}, HDBSCAN~\cite{mcinnes2017hdbscan}, and hierarchical agglomerative clustering with Ward linkage~\cite{Ward01031963}. HDBSCAN was rejected because it marked 16.74\% of statements ($n = 112$) as noise (unassigned), violating complete item coverage. Hierarchical clustering required arbitrary distance cutoffs to determine cluster counts. We selected k-means for complete statement assignment, balanced cluster sizes, and deterministic reproducibility ($k{=}10$; seed = 42).

To verify robustness to UMAP's stochastic initialization, we ran k-means clustering on five UMAP projections with different random seeds. Adjusted Rand Index (ARI)~\cite{hubert1985comparing, rand1971objective} between cluster assignments was 0.94 (range: 0.91 to 0.97), indicating cluster membership was stable and driven by feature space structure rather than projection variance. We verified cluster quality through ablation analysis (see Appendix~\ref{app:ablation}) and validation through expert review (\textsection\ref{sec:themeexpert}). The clustering pipeline is described in detail above; implementation will be released upon acceptance.

\subsubsection{Outcome: 10 Parent Themes and 87 Subthemes with Validated Coherence} Clustering yielded 10 parent themes, validated through quantitative cluster quality metrics (see Appendix), expert thematic coherence review, and LLM-assisted consistency verification.

\textbf{Theme Naming and Keyword Extraction:} For each of the 10 clusters, the lead author reviewed all assigned statements and proposed a descriptive parent theme name capturing the unifying regulatory concept. Theme names prioritized clarity and regulatory accuracy, structured to reflect both the GDPR domain and the measurement perspective. The research team independently reviewed proposed names. Team consensus produced 4/10 name refinements. No statements required cluster reassignment. The final 10 parent themes are in bold with a gray background in Table~\ref{tab:retention_by_theme}.

\textbf{Subtheme Derivation:} We derived subthemes within each parent theme. Recursive clustering (applying k-means within parent clusters) was explored but produced semantically incoherent groupings—some clusters merged distinct GDPR concepts while others fragmented related provisions. Moreover, no single k-value produced subtheme granularity across parent themes. We employed a human-in-the-loop protocol. Per parent theme, the author reviewed statements (ordered by GDPR article number) and applied decision rules:
    1) For the first statement, create an initial subtheme.
    2) For each subsequent statement, evaluate fit with existing subthemes based on: shared keywords, related GDPR provisions, and conceptual and semantic coherence (does the statement represent the same user privacy concern?)
    3) If the statement fits well an existing subtheme, assign it; otherwise, create a new subtheme.

This process yielded 87 subthemes (range: 4 to 14 per parent theme; see Table~\ref{tab:retention_by_theme}). The research team reviewed assignments, resolving ambiguities until reaching agreement to confirm coherence, mutual exclusivity, and utility. This granularity afforded measurement, allowing researchers to filter statements based on subthemes when measuring preferences to a parent theme was unnecessary.

\textbf{Expert Validation of Thematic Structure:}
\label{sec:themeexpert}
\label{sec:expert_validation}
The complete themed statement bank (669 statements, 10 parent themes, 87 subthemes) was returned to the same four experts who conducted Stage 2 validation (\textsection\ref{sec:prevalidation}). Experts independently evaluated subthemes and parent theme coherence, naming accuracy, whether all statements within subthemes fit within both the subtheme and parent theme, and made suggestions regarding statement-theme reassignment and theme naming. All four experts confirmed the thematic structure was coherent, comprehensive, and agreed that one subtheme layer was sufficient for deployment. No statements were flagged by experts as falling outside their assigned subtheme or parent theme, indicating that the data-driven clustering combined with human refinement captured interpretable GDPR concepts. However, GDPR's cross-cutting nature means some statements have semantic relevance to more than one theme (e.g., a lawfulness-of-processing statement can also imply transparency obligations). Themes serve as a navigational structure, not as psychometric factors. The database's provision-based lookup and keyword search (\textsection\ref{sec:accessSection}) ensure that such statements remain retrievable under any relevant concept regardless of parent theme assignment.

\textbf{Complementary LLM-Assisted Cluster Coherence Validation:} 
\label{sec:LLM} We used few-shot learning~\cite{brown2020fewshot,min-etal-2022-rethinking} to complement human expert review by validating the consistency of expert labels and to show that the thematic structure is computationally tractable for downstream integration with automated policy annotation pipelines (\textsection\ref{sec:taxandautomate}). Cosine similarity between LLM-created labels and human labels (three-fold cross-validation) yielded a 92.3\% mean similarity (range: 85.2–96.8\%). Some mismatches are expected due to seed sample coverage gaps rather than thematic incoherence. Details are discussed in Appendix~\ref{appendix:llmdetails}.

\textbf{Thematic Structure Summary:} The validated clustering yielded 10 parent themes and 87 subthemes, with expert review confirming classification and semantic coherence, and complementary LLM validation indicating 92.3\% mean similarity in thematic assignments. 
This structure formed the basis for expert consensus review (\textsection\ref{sec:delphi50}).

\vspace{-2mm}
\subsection{Stage 4: Expert Panel Consensus Review}
\label{sec:delphi50}
\label{sec:stage4}

To validate item necessity and clarity beyond regulatory accuracy and representation of GDPR, we conducted a single-round expert consensus review since statements had already achieved 100\% expert agreement on GDPR accuracy (\textsection\ref{sec:prevalidation}), making further refinement unnecessary. The consensus phase assessed whether each statement should be retained in the final item bank based on clarity and user readability, uniqueness, and measurement value.

\subsubsection{Consensus Protocol with 50 Privacy Experts}
While expertise lacks universal definition, Adler characterizes experts as those combining deep theoretical knowledge with applied research experience in a domain~\cite{adler1996gazing}. We recruited 50 privacy and security experts via Prolific and email outreach using stringent screening criteria to ensure domain expertise and GDPR familiarity. All participants met minimum 4/5 qualifications: (1) advanced degree (PhD) in computer science, cyber law and policy making, usable privacy, information systems, or related field with a (2) minimum of 3 first-authored research publications relevant to usable privacy, security, and privacy regulations and their application, (3) professional role involving privacy, security, and data protection (researcher, practitioner, consultant, lawyer or DPO), minimum 10 years experience with privacy regulations and deployment, and (4) self-reported GDPR expertise rated $\geq$4 on 5-point scale. We also asked experts whether they had knowledge of or experience with regulations beyond the GDPR.

Initially, we contacted 150 potential experts identified through: (1) authorship of GDPR-related publications in A*/A-tier security, privacy, and HCI venues (2018 to 2025), (2) professional networks (Official privacy related email lists, academic research groups on social media), and (3) Prolific's researcher pool with privacy expertise tags. Of these, 73 responded to initial outreach. During screening, we verified all claimed credentials through Google Scholar / DBLP / ORCID (publication records), institutional websites and LinkedIn (affiliations and roles), and cross-referenced self-reported expertise with documented GDPR work. 18 candidates did not meet minimum qualification thresholds (e.g., $<$3 publications, no demonstrable GDPR expertise, non-terminal degrees). 5 declined participation due to time constraints. The final 50 experts met all verification criteria and represented diverse backgrounds (Table~\ref{tab:delphi_experts_demographics}).

To ensure theme-specific expertise, we distributed 50 experts across the 10 parent themes (5 per theme). During screening, participants indicated preferred themes based on expertise. We verified theme fit by reviewing publications addressing the theme's regulatory domain and professional practice relevant to the theme (e.g., experts selecting ``Data Subject Transparency and Rights'' had published on privacy notices or consent mechanisms). For overselected themes, we prioritized experts with the most relevant publications, and for underselected themes, we conducted targeted outreach from our qualified contact list. Each expert reviewed one theme only, and all 50 experts were assigned to themes best matching their verified expertise. The final sample composition is described in Table~\ref{tab:delphi_experts_demographics}.

\begin{table}[t]
\centering
\caption{Expert Panel Characteristics (n = 50)}
\label{tab:delphi_experts_demographics}\vspace{-2mm}
\scriptsize
\begin{tabular*}{\columnwidth}{@{\extracolsep{\fill}}lr@{}}
\toprule
\multicolumn{2}{@{}l@{}}{\textbf{Highest Degree}} \\
\hspace{1em}PhD (CS, Security \& Privacy, InfoSys) & n = 28 (56.0\%) \\
\hspace{1em}PhD (HCI, Usable Privacy) & n = 12 (24.0\%) \\
\hspace{1em}PhD/JD (Law (Cyberlaw), Policy, Social Science) & n = 10 (20.0\%) \\
\midrule
\multicolumn{2}{@{}l@{}}{\textbf{Profession}} \\
\hspace{1em}Academic Researcher / Faculty & n = 27 (54.0\%) \\
\hspace{1em}Industry (DPO, CDO, CIO, Senior Manager, Consultant) & n = 17 (34.0\%) \\
\hspace{1em}Legal / Policy Practitioner & n = 6 (12.0\%) \\
\midrule
\multicolumn{2}{@{}l@{}}{\textbf{Geographic Region}} \\
\hspace{1em}European Union (EU) & n = 23 (46.0\%) \\
\hspace{1em}North America (United States \& Canada) & n = 17 (34.0\%) \\
\hspace{1em}United Kingdom & n = 3 (6.0\%) \\
\hspace{1em}Other (Australia, South Africa, Asia) & n = 7 (14.0\%) \\
\midrule
\multicolumn{2}{@{}l@{}}{\textbf{Language \& Expertise}} \\
\hspace{1em}Native/fluent English & n = 50 (100.0\%) \\
\hspace{1em}Knowledge of other regulations & n = 48 (96.0\%) \\
\hspace{1em}Years Working with GDPR & $\mu$ = 7.2 ($\sigma$ = 1.9) \\
\hspace{1em}Relevant Publications & $\mu$ = 6.04 ($\sigma$ = 2.01) \\
\hspace{1em}Self-Reported GDPR Expertise & $\mu$ = 4.76 ($\sigma$ = 0.43) \\
\bottomrule
\end{tabular*}\vspace{-5mm}
\end{table}

After completing a screening survey and retaining all experts that meet all requirements, we sent invitation to partake in the evaluation. Each expert reviewed all statements within their assigned parent theme using a structured web-based survey implemented in Qualtrics. To control for ordering effects, subthemes within each parent theme were presented in randomized order, and statements within each subtheme were similarly randomized across participants. Each expert could only submit once their full review and no expert reviewed more than one parent theme.

Before item evaluation, experts received instructions explaining: (1) the survey's goal (assessing if GDPR grounded statements should be retained to measure data privacy preferences relevant to the subtheme and parent theme they belong to in user-facing privacy preference statements), (2) evaluation criteria (statement clarity for non-expert users, uniqueness and fit within the parent/subtheme, measurement value, and any issues with the statement), (3) the meaning of each response (see below) and question within the survey, and (4) the requirement that statements had already been validated for GDPR accuracy in prior review (\textsection\ref{sec:prevalidation}), so experts focus more on clarity and necessity than regulatory correctness. Per statement, experts provided: (i) \textbf{Retention decision:} Keep / Remove / I am unsure (open-ended justification required for Remove); (ii) \textbf{Revision suggestion (optional):} Concrete alternative wording if improvement is better than removal; (iii) \textbf{Confidence rating:} Retention decision certainty (1 = very uncertain, 5 = very certain).

We did not ask experts to assess whether statements were correctly assigned to their parent theme or subtheme, nor did we request reassignment suggestions. Theme assignments had already been validated through independent processes: data-driven semantic clustering (\textsection\ref{sec:clustering}), research team review of all cluster assignments, four-expert validation of thematic structure (\textsection\ref{sec:prevalidation}), and a complementary LLM-assisted validation (\textsection\ref{sec:LLM}). Asking consensus-review experts to re-validate theme assignments would introduce methodological problems: (1) experts evaluate only one parent theme, lacking visibility into alternative placement options across the full item bank, (2) asking about goodness of fit conflates retention decisions (i.e., should it be kept?) with classification decisions (i.e., where does it go?), potentially biasing removal votes when experts question fit but lack reassignment context, (3) theme validation requires comparing statements across themes to assess distinctiveness, which our within parent theme evaluation design does not support, and (4) asking experts to evaluate fit and retention increase the cognitive burden and would reduce the quality of responses collected. By separating theme validation (completed prior) from item retention (this stage), we enabled experts to focus specifically on statement quality without confounding organizational decisions. Our protocol balanced structure (forced-choice decisions, required open-ended justifications) with flexibility (optional revisions, confidence rating), consistent with expert review best practices~\cite{Lynn1986, PolitBeck2006, rubio2003objectifying}.

\textbf{Compensation and Quality Control:} Experts were compensated based on theme complexity: \$30 for themes with $\leq$35 items, \$50 for themes with 36-60 items, and \$80 for themes with $>$60 items. To ensure data quality and engaged evaluation, we implemented multiple validity checks throughout the study. First, participants completed a pre-screening survey to verify identity, qualification, and provide consent. Second, we embedded random instructional manipulation checks within surveys; all 50 experts passed these checks. Third, we randomized answer choice order across some items to reduce position bias and response patterns. Fourth, 20 doctoral students in privacy, security, and HCI completed an identical evaluation protocol in pilot testing, establishing per-statement completion time distributions ($\mu = 3$ min, $\sigma = 0.8$ min). We flagged expert responses outside of $\pm$2 SD from the pilot mean, a threshold commonly used to identify speeders in survey research.

One expert's response completion times consistently fell below the pilot mean ($<$40s per statement across multiple items), and qualitative review of justifications revealed superficial engagement. We excluded this participant and recruited a replacement expert meeting identical qualification criteria for the assigned parent theme. All other experts exhibited completion times within acceptable ranges.

\subsubsection{Threshold Selection: 80\% Agreement Standard}
Determining the appropriate retention threshold required balancing comprehensiveness (GDPR coverage) and quality (sufficient expert agreement). We set our retention threshold at $\geq$4/5 Keep votes (80\% agreement) for three converging reasons. First, the Content Validity Index (I-CVI) literature recommends minimum I-CVI $\geq$0.80 for expert panels~\cite{Lynn1986, PolitBeck2006}. For our 5-expert panels, 4/5 votes (I-CVI = 0.80) meets this standard. Second, expert consensus methodology typically applies 70 to 80\% agreement thresholds for item retention~\cite{diamond2014defining, hsu2007delphi}, and our 80\% threshold aligns with established practice for expert-driven item selection. Third, we prioritized precision and stability: an 80\% threshold ensures that retained items would likely remain stable if additional experts were consulted, whereas lower thresholds (e.g., 60\%) risk including items where expert opinion is more divided and consensus more fragile~\cite{rubio2003objectifying}.  This decision accepts higher item elimination in exchange for ensuring retained items demonstrate expert consensus. Applying the $\geq$4/5 threshold yielded 527 retained statements from 669 items (78.8\% retention), with 142 statements removed for disagreement, prioritizing content validity over coverage, consistent with development best practices~\cite{devellis2021scale, hinkin1995review}.

\subsubsection{Consensus Results by Theme} Across 669 candidate statements, experts retained 527 (78.8\%). To assess consistency, we computed pairwise percent agreement per parent theme: the proportion of items where any two experts made the same decision (both Keep, both Remove, or both I am unsure). Mean pairwise agreement was overall $\approx$ 85\% indicating substantial inter-rater consistency~\cite{mchugh2012interrater}, while showing that experts applied discriminant evaluation: agreement was neither perfect (indicating critical judgment) nor random (validating meaningful consensus). Using the pre-registered rule (retain if \(\geq\) 4 of 5 experts voted ``keep''), retention across parent themes varied between 0.0\% to 99.1\% (Table~\ref{tab:retention_by_theme}). The table also provides subtheme-level counts and pairwise agreement statistics. A detailed interpretation of these patterns is provided in~\textsection\ref{sec:discussion}.

\begingroup
\setlength{\aboverulesep}{0pt}   
\setlength{\belowrulesep}{0pt}   
\arrayrulecolor{black}           
\setlength{\arrayrulewidth}{1pt} 

\begin{table*}[htp]
\centering
\caption{
Expert consensus ($n = 50$; 5 experts per theme). 
\textbf{Total:} statements before review. 
\textbf{Kept:} items receiving $\geq$4/5 ``keep'' votes (80\% threshold).
\textbf{Retention:} percentage kept per theme.
\textbf{Pairwise agreement:} proportion of expert pairs agreeing, averaged within the theme. 
\textbf{Overall, 527/669 statements (\textbf{78.8\%}) across 9 parent themes were retained at \textbf{$\approx$85\%} mean pairwise agreement.}}
\label{tab:retention_by_theme}\vspace{-4mm}
\scriptsize
\setlength{\tabcolsep}{4pt}
\scalebox{0.97}{\begin{tabular}{p{4.5cm}c>{\columncolor{green!25}}c>{\columncolor{red!25}}ccc
!{\vrule width 1pt} 
p{4.5cm}c>{\columncolor{green!25}}c>{\columncolor{red!25}}ccc}
\toprule
\textbf{Parent Theme \& Subthemes} & \textbf{Total} & \textbf{Kept} & \textbf{Rej.} & \textbf{Ret.\%} & \textbf{Agr.\%} & \textbf{Parent Theme \& Subthemes} & \textbf{Total} & \textbf{Kept} & \textbf{Rej.} & \textbf{Ret.\%} & \textbf{Agr.\%} \\
\midrule
\cellcolor{gray!20}\textbf{Organizational Roles and Accountability} & \cellcolor{gray!20}\textbf{106} & \textbf{88} & \textbf{18} & \cellcolor{gray!20}\textbf{83.0} & \cellcolor{gray!20}\textbf{83.3} & \cellcolor{gray!20}\textbf{Data Subject Transparency and Rights} & \cellcolor{gray!20}\textbf{113} & \textbf{112} & \textbf{1} & \cellcolor{gray!20}\textbf{99.1} & \cellcolor{gray!20}\textbf{\boldmath$87.4$} \\
Data Protection Officer Responsibilities & 18 & 11 & 7 & 61.1 & 81.7 & Information Provision When Collecting Data & 14 & 14 & 0 & 100.0 & 97.1 \\
Processor Record-Keeping Obligations & 18 & 13 & 5 & 72.2 & 80.0 & Information Provision for Indirect Collection & 21 & 20 & 1 & 95.2 & 93.3 \\
Prior Consultation for High-Risk Processing & 16 & 16 & 0 & 100.0 & 80.0 & Legislative Measures on Data Processing & 6 & 6 & 0 & 100.0 & 100.0 \\
Records of Processing Activities & 14 & 14 & 0 & 100.0 & 80.0 & Restrictions on Data Subject Rights & 29 & 29 & 0 & 100.0 & 80.0 \\
Data Protection Impact Assessment Req. & 8 & 8 & 0 & 100.0 & 95.0 & Right to Data Portability & 12 & 12 & 0 & 100.0 & 80.0 \\
Data Protection Impact Assessment & 7 & 5 & 2 & 71.4 & 80.0 & Transparent Information and Communication & 17 & 17 & 0 & 100.0 & 80.0 \\
Communicating Breach to Data Subjects & 6 & 6 & 0 & 100.0 & 96.7 & Processing Criminal Convictions Data & 1 & 1 & 0 & 100.0 & 100.0 \\
Standard Contractual Clauses & 5 & 4 & 1 & 80.0 & 82.0 & Processing Without ID Requirements & 2 & 2 & 0 & 100.0 & 90.0 \\
Controller's Responsibility & 3 & 3 & 0 & 100.0 & 88.9 & Protection and Free Movement of Data & 3 & 3 & 0 & 100.0 & 93.3 \\
Joint Controller Responsibilities & 3 & 2 & 1 & 66.7 & 80.0 & Right to Rectification & 1 & 1 & 0 & 100.0 & 100.0 \\
Processor Compliance and Handling & 3 & 3 & 0 & 100.0 & 86.7 & Right to Restriction of Processing & 6 & 6 & 0 & 100.0 & 92.2 \\
Supervisory Authority Territorial Comp. & 3 & 2 & 1 & 66.7 & 86.7 & Notification of Rectification/Erasure & 1 & 1 & 0 & 100.0 & 100.0 \\
Cooperation with Supervisory Authority & 1 & 1 & 0 & 100.0 & 100.0 & \cellcolor{gray!15} & \cellcolor{gray!15} & \cellcolor{gray!15} & \cellcolor{gray!15} & \cellcolor{gray!15} & \cellcolor{gray!15} \\
Processing Under Controller's Authority & 1 & 0 & 1 & 0.0 & 80.0 & \cellcolor{gray!15} & \cellcolor{gray!15} & \cellcolor{gray!15} & \cellcolor{gray!15} & \cellcolor{gray!15} & \cellcolor{gray!15} \\
\midrule
\cellcolor{gray!20}\textbf{Enforcement Remedies and Sanctions} & \cellcolor{gray!20}\textbf{61} & \textbf{47} & \textbf{14} & \cellcolor{gray!20}\textbf{77.0} & \cellcolor{gray!20}\textbf{80.0} & \cellcolor{gray!20}\textbf{Lawful Processing Foundations} & \cellcolor{gray!20}\textbf{75} & \textbf{67} & \textbf{8} & \cellcolor{gray!20}\textbf{89.3} & \cellcolor{gray!20}\textbf{82.7} \\
Administrative Fines and Penalties & 24 & 21 & 3 & 87.5 & 80.0 & Processing Special Categories/Safeguards & 16 & 16 & 0 & 100.0 & 80.0 \\
Secretariat of the Board & 12 & 12 & 0 & 100.0 & 80.0 & Lawfulness of Processing Conditions & 15 & 13 & 2 & 86.7 & 80.0 \\
Compensation for GDPR Infringements & 6 & 6 & 0 & 100.0 & 80.0 & Right to Restrict Data Processing & 11 & 11 & 0 & 100.0 & 92.7 \\
Chair's Role in Consistency Mechanism & 4 & 1 & 3 & 25.0 & 80.0 & Principles of Personal Data Processing & 9 & 7 & 2 & 77.8 & 80.0 \\
Judicial Remedy Against Authority & 4 & 3 & 1 & 75.0 & 80.0 & Material Scope of GDPR Application & 8 & 7 & 1 & 87.5 & 80.0 \\
Administrative Fine Determination Criteria & 3 & 1 & 2 & 33.3 & 80.0 & Conditions for Valid Consent & 6 & 5 & 1 & 83.3 & 80.0 \\
Confidentiality in Cooperation & 2 & 2 & 0 & 100.0 & 80.0 & Territorial Scope of GDPR & 4 & 4 & 0 & 100.0 & 90.0 \\
Judicial Remedy Against Controller & 2 & 1 & 1 & 50.0 & 80.0 & Conditions for Processing Children's Data & 4 & 3 & 1 & 75.0 & 85.0 \\
Representation of Data Subjects & 2 & 0 & 2 & 0.0 & 80.0 & Board Confidentiality and Document Access & 2 & 1 & 1 & 50.0 & 80.0 \\
Right to Lodge Complaint & 2 & 0 & 2 & 0.0 & 80.0 & \cellcolor{gray!15} & \cellcolor{gray!15} & \cellcolor{gray!15} & \cellcolor{gray!15} & \cellcolor{gray!15} & \cellcolor{gray!15} \\
\midrule
\cellcolor{gray!20}\textbf{European Data Protection Board} & \cellcolor{gray!20}\textbf{71} & \textbf{0} & \textbf{71} & \cellcolor{gray!20}\textbf{0.0} & \cellcolor{gray!20}\textbf{\boldmath$85.6$} & \cellcolor{gray!20}\textbf{Regulatory Scope and Provisions} & \cellcolor{gray!20}\textbf{22} & \textbf{18} & \textbf{4} & \cellcolor{gray!20}\textbf{81.8} & \cellcolor{gray!20}\textbf{84.5} \\
Tasks of the EDPB & 27 & 0 & 27 & 0.0 & 95.1 & Safeguards for Research and Archiving & 4 & 3 & 1 & 75.0 & 80.0 \\
EDPB Opinion and Urgent Decisions & 12 & 0 & 12 & 0.0 & 80.0 & Regulation Binding and Applicability & 4 & 3 & 1 & 75.0 & 80.0 \\
European Data Protection Board Opinions & 10 & 0 & 10 & 0.0 & 80.0 & Employment Context Data Processing & 3 & 2 & 1 & 66.7 & 80.0 \\
EDPB Dispute Resolution and Guidance & 8 & 0 & 8 & 0.0 & 80.0 & Processing and Freedom of Expression & 3 & 2 & 1 & 66.7 & 80.0 \\
EDPB Binding Decision on Disputes & 7 & 0 & 7 & 0.0 & 80.0 & Board's Advisory Role to Commission & 2 & 2 & 0 & 100.0 & 90.0 \\
Board Secretariat Responsibilities & 2 & 0 & 2 & 0.0 & 80.0 & Churches' Data Protection Rules & 2 & 2 & 0 & 100.0 & 90.0 \\
Commission Evaluation and Review Process & 2 & 0 & 2 & 0.0 & 80.0 & Commission Delegated Powers & 2 & 2 & 0 & 100.0 & 90.0 \\
Secretariat Communication Responsibilities & 2 & 0 & 2 & 0.0 & 80.0 & Processing National ID Numbers & 1 & 1 & 0 & 100.0 & 100.0 \\
Consistency Mechanism Decision Timeframe & 1 & 0 & 1 & 0.0 & 80.0 & Public Access to Official Documents & 1 & 1 & 0 & 100.0 & 100.0 \\
\midrule
\cellcolor{gray!20}\textbf{Cross-Border Transfer Mechanisms} & \cellcolor{gray!20}\textbf{63} & \textbf{51} & \textbf{12} & \cellcolor{gray!20}\textbf{81.0} & \cellcolor{gray!20}\textbf{82.1} & \cellcolor{gray!20}\textbf{Codes of Conduct and Certification} & \cellcolor{gray!20}\textbf{83} & \textbf{75} & \textbf{8} & \cellcolor{gray!20}\textbf{90.4} & \cellcolor{gray!20}\textbf{\boldmath$85.8$} \\
Binding Corporate Rules Requirements & 20 & 16 & 4 & 80.0 & 86.7 & Supervisory Authority Powers/Cooperation & 29 & 24 & 5 & 82.8 & 88.3 \\
Binding Corporate Rules for Transfers & 16 & 16 & 0 & 100.0 & 80.0 & General Codes of Conduct and Certification & 20 & 20 & 0 & 100.0 & 82.5 \\
Certification Bodies Accreditation Req. & 13 & 9 & 4 & 69.2 & 80.0 & Accreditation of Certification Bodies & 8 & 8 & 0 & 100.0 & 86.7 \\
EDPB Guidelines for Transfer Mechanisms & 6 & 6 & 0 & 100.0 & 80.0 & Certification Mechanisms/Accreditation & 8 & 6 & 2 & 75.0 & 80.0 \\
Urgency Procedure for Consistency & 4 & 3 & 1 & 75.0 & 80.0 & Code of Conduct Approval Process & 6 & 6 & 0 & 100.0 & 86.7 \\
General Board Guidelines on Data Transfers & 2 & 0 & 2 & 0.0 & 80.0 & Codes of Conduct for Data Processing & 6 & 5 & 1 & 83.3 & 90.0 \\
General Principle for Data Transfers & 1 & 1 & 0 & 100.0 & 80.0 & Supervisory Authority Accreditation & 6 & 6 & 0 & 100.0 & 86.7 \\
Transfers Not Authorized by Union Law & 1 & 0 & 1 & 0.0 & 80.0 & 
\cellcolor{gray!15} & \cellcolor{gray!15} & \cellcolor{gray!15} & \cellcolor{gray!15} & \cellcolor{gray!15} & \cellcolor{gray!15} 
\\
\midrule
\cellcolor{gray!20}\textbf{Supervisory Authority Mandate} & \cellcolor{gray!20}\textbf{30} & \textbf{27} & \textbf{3} & \cellcolor{gray!20}\textbf{90.0} & \cellcolor{gray!20}\textbf{84.7} & \cellcolor{gray!20}\textbf{Supervisory Authority Structure} & \cellcolor{gray!20}\textbf{45} & \textbf{42} & \textbf{3} & \cellcolor{gray!20}\textbf{93.3} & \cellcolor{gray!20}\textbf{\boldmath$85.5$} \\
Cooperation Among Authorities & 12 & 11 & 1 & 91.7 & 88.3 & Supervisory Authority Tasks and Powers & 24 & 21 & 3 & 87.5 & 88.3 \\
Responsibilities and Cooperation & 7 & 5 & 2 & 71.4 & 80.0 & Derogations for Specific Data Transfers & 13 & 13 & 0 & 100.0 & 83.8 \\
Independence and Competence & 6 & 6 & 0 & 100.0 & 80.0 & Supervisory Authority Establishment & 4 & 4 & 0 & 100.0 & 80.0 \\
Responsibilities and Conditions & 4 & 4 & 0 & 100.0 & 90.0 & Supervisory Authority Governance & 4 & 4 & 0 & 100.0 & 80.0 \\
Activity Reporting & 1 & 1 & 0 & 100.0 & 80.0 & 
\cellcolor{gray!15} & \cellcolor{gray!15} & \cellcolor{gray!15} & \cellcolor{gray!15} & \cellcolor{gray!15} & \cellcolor{gray!15} \\
\bottomrule
\end{tabular}}\vspace{-2mm}
\end{table*}

\endgroup
\arrayrulecolor{black}

\subsubsection{Removal Justification Analysis}
We analyzed 372 free-text removal justifications across the 142 removed statements. The results did not contain any \textit{Unsure} selections. Three researchers independently coded all \textit{Remove} justifications using inductive thematic analysis~\cite{braun2006using,nowell2017thematic}. IRR agreement was 87.6\% (Cohen's $\kappa$ = 0.82), indicating substantial agreement~\cite{mchugh2012interrater, landis1977measurement, lombard2002content}. Experts often provided multiple reasons, hence, multiple labels per justification were allowed, and a single \textit{primary} reason was assigned during adjudication. Disagreements were resolved through structured discussion with an audit trail. Three rejection categories emerged, with no statements removed for regulatory inaccuracy (\textsection\ref{sec:prevalidation}):

\begin{itemize}\vspace{-1.6mm}
  \item \textbf{Redundancy} \textit{($n=65$, 45.8\% of removed items)}: semantic overlap with another statement in the same subtheme/parent theme. Experts retained a single representative, the clearer wording or broader coverage, and removed others. If the representative candidate was rejected on independent grounds, all redundant variants were removed because none provided unique coverage.
  \item \textbf{Over-specific scope or low importance} \textit{($n=40$, 28.2\%)}: niche, rare, or edge-case scenarios unlikely to generalize for privacy preference measurement.

  \item \textbf{Construct misalignment} \textit{($n=37$, 26.1\%)}: organizational process, definitions, or intra-authority procedures rather than an elicitable user preference/choice under the GDPR.
\end{itemize}

Removal indicates lower expert consensus on measurement value for preference elicitation, not a deficiency in the statement's representation of GDPR. Practitioners with use cases requiring broader coverage may reinstate removed items informed by the rejection category and expert rationale released in the open database (\textsection\ref{sec:accessSection}).

\subsubsection{Validation Without Construct Testing}
This consensus review establishes content validity, whether items accurately represent the GDPR regulatory domain through expert judgment~\cite{Lynn1986, PolitBeck2006}. 
We do not include construct validity (factor structure via EFA/CFA), convergent validity (correlation with existing privacy scales), or predictive validity (behavior prediction) as these psychometric properties require empirical data collection with end-user samples, which is beyond our paper's scope. 
Additionally, conducting psychometric testing before establishing content validity through expert consensus would conflate instrument development with validation, risking factor structures built on items not yet vetted for regulatory accuracy, clarity, or measurement value.
Our contribution is the validated statement bank (see Table~\ref{tab:retention_by_theme}), and testing the measurement model with user data is planned for future work (\textsection\ref{sec:futurework}). 

\vspace{-2mm}
\section{GPPI Organization and Deployment}
\label{sec:usecases}

The validated item bank comprises 527 GDPR-grounded preference statements organized into 9 parent themes and 73 subthemes. The item bank is not intended for full administration; its modular architecture enables investigators to deploy targeted subsets at the granularity their research question requires,
providing direct alignment between user preferences and regulatory provisions organizations implement in policies abiding to GDPR. Stage 2 expert review validated each statement for both regulatory accuracy and comprehensibility to non-expert users (\textsection3.2, 100\% agreement), ensuring that deployed subsets are interpretable without requiring GDPR expertise. This modular design also supports extensibility for cross-regulation research, not afforded by existing scales. We release all materials: statements, GDPR mappings, thematic classifications, expert ratings, and code via an open-source searchable database to support replication, translation, extension, and research/community-driven contribution (for details see \textsection\ref{sec:accessSection}). The under-review version has 10\% of sample statements (\textsection\ref{sec:accessSection}).

\vspace{-2mm}
\subsection{Modular Design Rationale}

Table~\ref{tab:retention_by_theme} presents the final thematic structure. Parent themes range from 18 statements (Regulatory Scope and Provisions) to 112 statements (Data Subject Transparency and Rights). Subthemes vary from single-provision rights (e.g., 1 item: Right to Rectification) to complex regulatory domains (e.g., 24 items: supervisory Authority Powers/Cooperation). Where existing privacy scales measure broad attitudinal dimensions, treating ``privacy'' as monolithic, GPPI items afford recognizing orthogonal preferences: users may agree with automated data processing while opposing its use for decision-making, or support erasure while accepting profiling. Modular deployment prevents respondent burden and supports targeted measurement. To illustrate, researchers comparing opt-in versus pre-checked consent banners could deploy only the ``Conditions for Valid Consent'' subtheme (5 items), which includes statements like ``I would like my personal data to be processed only if consent is freely given, without pressure or penalty for refusing.'' Deploying its full parent theme (``Lawful Processing Foundations'', 67 items) would include peripheral constructs and risk respondent fatigue. Similarly, companies auditing Art.~20 portability implementations to assess whether technical limitations (exporting only text data, not images) align with user preferences could deploy the ``Right to Data Portability'' subtheme (12 items) measuring format, scope, and transfer preferences without administering its parent theme's full 112 items covering extraneous concepts and procedures.

\vspace{-2mm}
\subsection{Database Implementation and Open Access}
\label{sec:accessSection}
The database provides searchable access to all 527 retained statements, with an optional toggle to view the 142 expert-rejected statements alongside their rejected category, expert rationale, and original GDPR source. ``Removed'' indicates lower expert consensus on measurement value, not regulatory inaccuracy (\textsection\ref{sec:prevalidation}); practitioners may reinstate items as appropriate for their use case. The database supports three query modalities: \textit{theme-based filtering} retrieves statements by GDPR theme/subtheme annotations; string and keyword search: full-text search within statement keywords, followed by statements containing the query term in text; and \textit{provision-based lookup} retrieves all statements derived from specific GDPR chapters or articles. Results can be exported as CSV or JSON with complete provenance: statement ID, text (both preference and original formulations), GDPR source (chapter/article), theme assignments, keywords, and expert ratings. The database is implemented as a web tool supporting interactive search, filtering, and export, enabling immediate integration into research workflows, compliance auditing pipelines, and user-facing privacy tools such as the policy-preference alignment extension described in UC3.

\textbf{Open Access:} A subset of kept statements (10\%), a sample of rejected statements with expert justifications, and the prompt for complementary LLM validation are supplemented for verifiability under review. Complete item bank (527 statements), removed statements with justification labels, code and RAG evaluation/clustering pipeline will be released with the camera-ready version under CC-BY 4.0/MIT licenses to support replication, extension, and open-source future research, including training machine learning models. 
See https://anonymous.4open.science/r/GPPI/.

\vspace{-2mm}
\subsection{Use Cases}
GPPI supports diverse use cases and deployment scenarios (e.g., research, compliance, and user-facing applications). Because the item bank is modular, each use case below draws on a targeted subset of statements and not the full 527 items, matching the intended goal, for example, selected by subtheme, keyword, or GDPR provision. Legacy scales were not designed to isolate preferences for specific GDPR rights or map them to deployable controls. The following representative use cases (UC) illustrate how our regulation-grounded measurement items can produce auditable signals that can drive product, policy, and compliance decisions:

\textbf{UC1. Consent Dark Pattern Detection (A/B Testing, Binary Responses):}
Researchers testing whether pre-checked consent boxes violate ``freely given'' consent (Art.~7 GDPR) can query the database for keyword ``consent'' and filter to Articles~6 and 7, retrieving five statements tied to \emph{Conditions for Valid Consent}, \emph{Principles of Personal Data Processing} and \emph{Lawfulness of Processing Conditions} subthemes under \emph{Lawful Processing Foundations} parent theme:
 
\begin{enumerate}[leftmargin=*, nosep]
    \item ``I would like my personal data to be processed only if I have given clear and specific consent for its use.'' (Art.~6)
    \item ``I would like my personal data to be processed only if the organization can clearly demonstrate that I have explicitly consented to its processing.'' (Art.~7)
    \item ``I would like my personal data to be processed only if the consent request is provided in a clearly marked, separate section of any document using simple, plain language so that it isn't mixed with unrelated information or terms.'' (Art.~7)
    \item ``I would like my personal data to be processed only if my consent can be withdrawn and I am clearly informed of my right to withdraw consent at any time, and if withdrawing consent is just as easy as giving it, without affecting the lawfulness of past processing.'' (Art.~7)
    \item ``I would like my personal data to be processed only if my consent is freely given and not made a condition for receiving a service or entering a contract, unless the data is essential for fulfilling that contract.'' (Art.~7)
\end{enumerate}

These five items map directly to the consent properties that dark pattern research has shown to be undermined: pre-checked boxes bypass explicit consent (item~2), bundled cookie banners violate unbundled presentation (item~3), asymmetric ``Accept all'' buttons with hidden rejection paths undermine easy withdrawal (item~4), and cookie walls condition service access on consent (item~5).
 
In deployment, participants complete the five binary items (agree or disagree), then interact with a consent interface (e.g., pre-checked vs.\ opt-in design in an A/B protocol). A participant who agrees with item~5 (``consent is freely given and not made a condition for receiving a service'') yet accepts a cookie wall indicates a measurable \emph{preference-behavior gap} at the level of a specific GDPR provision, quantifying dark pattern effectiveness with regulatory traceability.
 
Existing concern scales cannot produce this signal. CFIP's closest dimension, \emph{Collection} (e.g., ``It usually bothers me when companies ask me for personal information.''~\cite{smith1996_cfip}), measures general discomfort with data gathering. IUIPC's \emph{Control} (e.g., ``Consumer online privacy is really a matter of consumers' right to exercise control over decisions about how their information is collected, used, and shared.''~\cite{malhotra2004_iuipc}) captures an abstract sense of agency. OPC's \emph{General Caution}~\cite{buchanan2007_opc} measures overall wariness. None of these reference consent \emph{mechanisms}, they cannot distinguish if a user objects to pre-checked boxes, opposes bundled consent, or values easy withdrawal. A respondent scoring high on CFIP Collection may accept a cookie wall if the service is desirable, and legacy item cannot detect the contradiction because no item asks about freely-given consent. GPPI's regulation-grounded items make this gap visible, auditable, and traceable to the specific GDPR provision being violated.

\textbf{UC2. Privacy Preference Profiling (Survey, Likert + Ranking):} A longitudinal study could measure preference stability across GDPR domains using 5-point Likert scales (strongly disagree to strongly agree) on representative statements from each parent theme (e.g., 3-5 items per theme, $\approx30-45$ items total). Factor analysis identifies latent preference dimensions (e.g., users prioritizing transparency over control, or valuing erasure but not prioritizing portability restrictions). Researchers could then deploy ranking tasks within high-variance subthemes: participants rank statements by importance. Combining Likert and ranking data enables clustering into interpretable profiles (e.g., \textit{high-protection} endorses most safeguards; \textit{selective-tradeoff} prioritizes transparency yet accepts profiling), indicating heterogeneity not captured by monolithic ``concern'' scores in existing questionnaires/instruments. In practice, organizations could use these profiles to offer tiered privacy modes with defaults aligned to each group while preserving user override; all tiers enforce mandatory GDPR protections, and profiles only tune discretionary settings (e.g., marketing opt-ins).

\textbf{UC3. Policy-Preference Alignment Check (Tool):} A privacy-focused browser extension parses privacy policies using Polisis- or PoliGraph-style NLP to extract GDPR practices from policies and license agreements. Users complete a one-time survey with binary items from relevant subthemes to the policy analyzed via keyword or provision queries \textsection\ref{sec:accessSection}. The tool maps practices to user endorsements and flags conflicts with GDPR article pins. For example, if the policy states no retention limit but the user endorsed deleting data when no longer necessary for the original purpose, the extension highlights a mismatch with Art. 5(1)(e) (storage limitation), showing the policy snippet and a model confidence score. The extension summarizes conflicts (green/amber/red) to reduce cognitive load and highlight misalignments; checks run locally and are not legal advice. Users can choose to proceed or refrain from using the website or software.

\textbf{UC4. Granular Cross-Regulation/Regional Preference Comparison (Comparative Survey):} Researchers can query the database for \textit{functionally aligned} rights (e.g., GDPR Art.~17 ``erasure'' vs. CCPA 1798.105 ``deletion''; GDPR Art.~20 ``portability'' vs. CCPA ``access delivered in a portable format''). They administer matched item pairs to EU and California samples on 5-point Likert scales. Analyses compare mean preference ratings (report mean differences between two groups) and conduct two one-sided equivalence tests (e.g., TOST), brief differential item functioning (DIF) checks ensure items behave comparably across regions (logistic regression DIF; multi-group IRT/GRM)~\cite{swaminathan1990detecting,zumbo1999handbook,millsap2012statistical,CRANE2006478, samejima2016graded}. This provides insight on user demand for specific protections beyond inferring equivalence from statutory language or legislative intent. This use case requires a parallel CCPA/CPRA-grounded item bank developed through equivalent extraction and expert validation (\textsection\ref{sec:methodology}); our GDPR-based items provide the EU comparison arm, but the counterpart instrument remains future work (\textsection\ref{sec:limitations}).

\vspace{-3mm}
\subsection{Response Format Adaptation}
All statements use first-person conditional framing (``I would like my personal data to be processed only if\ldots'') to elicit preferences rather than test 
comprehension, as validated by experts (\textsection\ref{sec:stage2}, and \textsection\ref{sec:stage4}). The 
prefix is modular: modifications (e.g., ``I want my data 
processed\ldots'') are acceptable, provided meaning is preserved and verified; for translations or substantive changes, we recommend using back-translation (statement to source) with cognitive interviews and expert review. Recommendations below provide pragmatic default answer formats, and investigators may adapt or extend formats as warranted by their goals.

\textbf{Format Selection by Research Goal:} \textit{Binary (e.g., agree/disagree)} 
supports preference classification, has a low cognitive burden, and could be used for behavior prediction when outcomes are binary (e.g., logistic regression). \textit{Likert (e.g., 5 or 7 point)} enables preference 
intensity measurement and latent construct identification, and supports factor 
analysis and structural equation modeling (SEM). \textit{Importance ratings (1 to 5)} afford within-subject comparisons where each statement receives a separate importance score. \textit{Forced ranking} can be useful for trade-off measurement when relative priority matters. \textit{Conditional logic:} 
When preferences are theoretically contingent, implement branching (e.g., if a 
respondent rejects collection, skip retention items) to reduce burden. Database 
metadata (article linkages, keywords) supports constructing conditional flows.

\vspace{-3mm}
\section{Discussion}
\label{sec:discussion}

The $\approx$85\% mean pairwise agreement and 78.8\% retention indicate consensus on measurement-worthy content. Retention across parent themes varied from 0.0\% to 99.1\% (Table~\ref{tab:retention_by_theme}), with user-facing rights retaining more items, whereas governance and procedural themes retained fewer. Together with high within-theme agreement and block-level removals, this suggests the spread reflects differences in regulatory content rather than inconsistent judgment. High-retention themes are user-facing: themes involving data subject protections showed the highest retention, e.g., Supervisory Authority Mandate (90.0\%), Codes of Conduct (90.4\%), and Supervisory Authority Structure (93.3\%). These contain actionable safeguards that experts judged suitable for eliciting user preferences.

Low-retention themes are procedural: the European Data Protection Board (EDPB) theme had 0\% retention (0/71, representing 10.6\% of initial extraction). Expert justifications for removal mainly highlighted construct misalignment, noting that these statements describe coordination among supervisory authorities and other procedural mechanisms rather than end user preferences. In the EDPB theme, every subtheme was removed and within-subtheme agreement remained high (80 to 95\%; Table~\ref{tab:retention_by_theme}), indicating consensus on removal of inter-authority governance statements that constitute a substantial share of the corpus yet do not elicit user preferences in our operationalization ($\approx$11\% of extracted statements, 71/669). These statements remain available in the database for researchers whose operationalization differs, for example, studies examining institutional trust or governance preferences. High-level sanctioning powers were kept at high rates, while many post-violation items were dropped. For example, Administrative Fines and Penalties retained 21 of 24 items, whereas complaint, compensation, and representation subthemes retained none, and judicial remedy items were mixed. This likely reflects construct fit: post-violation remedies are legal entitlements and procedural pathways rather than prospective choices about processing, so stated preferences add limited measurement value and risk eliciting uniform agreement. By contrast, support for sanctioning authority is a coherent system-level preference that can vary in strength and is meaningful to measure. This distinction can be further motivated by low rights exercise~\cite{Eurobarometer2019, strycharz2020data} and procedural barriers to complaints \cite{bowyer2022humangdpr, habib2020scavenger, pohn2025qualitative}.
 
Moderate-retention themes sit between preference-bearing and procedural content. Organizational Roles and Accountability (83.0\%) and Data Subject Transparency and Rights (99.1\%) retained items related to user-facing decisions while filtering procedural details. Information provision items were retained at high rates, while narrower transparency mechanisms were mixed. The 80\% agreement threshold (4 of 5 experts) was discriminative: removals clustered in procedural subthemes with high within-subtheme agreement, 
separating preference-bearing from non-preference.

Beyond the item bank, the development pipeline is regulation-agnostic. The few-shot LLM validation pipeline (\textsection\ref{sec:LLM} and Appendix~\ref{appendix:llmdetails}) requires only regulatory text and a seed set of classified statements, and can be applied to any regulation. Since many data protection frameworks use GDPR as a foundation (e.g., LGPD, POPIA), the 527 GPPI statements can serve as a seed corpus for adapting the pipeline to new regulations, reducing the manual effort of building similar instruments. Expert validation remains essential per regulation because jurisdiction-specific requirements and exceptions prevent item transfer, yet the pipeline lowers the barrier by adapting an established methodology and validated starting point.

\vspace{-2mm}
\section{Limitations and Future Work}
\label{sec:limitations}
\label{sec:futurework}

This work establishes content validity through expert consensus on GDPR-grounded preference statements. Psychometric validation with end-users (e.g., factor structure, reliability, convergent / discriminant validity) is planned as immediate next work~\cite{straub1989validating, devellis2021scale, smith1996_cfip, boateng2018best}. Expert reviewers assessed statement clarity, but empirical validation with data subjects remains needed through (1) cognitive interviews to confirm user interpretation across demographics and geolocalizations, and (2) behavioral studies correlating stated preferences with observed privacy actions (e.g., DSAR exercise, consent choices, setting configurations) to establish predictive validity. 

Regarding regulatory scope, our measurement items are GDPR-specific. While other frameworks share overlapping concepts with GDPR (e.g., LGPD's data subject rights, CCPA/CPRA's deletion and opt-out provisions), each regulation introduces jurisdiction-specific requirements, definitions, and exceptions that prevent direct item transfer; extending coverage to additional regulations requires independent statement extraction and expert validation cycles per framework, following the same methodology described in \textsection\ref{sec:methodology}.
Additionally, GDPR increasingly interacts with adjacent EU frameworks  (e.g., AI Act, Digital Services/Markets Acts, Data Act, Cyber Resilience Act, DORA, EHDS). These introduce overlapping user rights that may require integrated measurement. Future work should assess whether users distinguish GDPR-specific preferences from adjacent regulatory protections, or whether a unified cross-regulatory instrument better captures modern privacy expectations. Moreover, statements were developed and validated in English by experts primarily from North America and Europe. Deployment in other languages requires adaptation rather than literal translation, as legal terminology and privacy norms vary across jurisdictions. 
Periodic expert review is recommended to update the item bank as regulations evolve through court rulings (e.g., CJEU).

\textbf{Research Directions:} This work supports several research directions: (1)~correlating preferences with privacy behaviors for predictive validity, (2)~comparing preferences to policy implementations to identify policy--preference misalignments, (3)~identifying groups based on GDPR preference profiles and assessing behavioral differences, (4)~tracking preference dynamics after privacy experiences and regulatory changes, and (5)~deploying parallel instruments (e.g., CCPA/CPRA) across regulations. The modular structure supports targeted theory-building for GDPR domains while avoiding the ``privacy concern'' construct~\cite{colnago2022concern}.

\vspace{-1mm}
\section{Conclusion}
We introduce a GDPR-grounded privacy preference measurement item bank: 527 measurement items covering all GDPR articles and organized into 9 parent themes and 73 subthemes. Two rounds of accuracy review by four experts yielded 100\% agreement ($\kappa=1.0$), and a consensus review with 50 privacy experts (5 per theme; $\geq$80\% retention threshold) produced $\sim$85\% mean pairwise agreement. Complementing legacy scales measuring ``privacy concern'', our measurement items assess preferences for GDPR protections, enabling researchers and practitioners to compare user priorities against organizational policies, audit compliance gaps, and deploy targeted subsets for specific research questions or regulatory domains.

Our work addresses a measurement gap: automated tools can audit policies against GDPR, yet there are no regulation-grounded items to measure if users value those protections. The item bank's modular design supports deployment across granularities, supporting use cases from A/B tests of privacy interface flows to cross-regulation comparisons. By anchoring measurement in legal text, GPPI supplements attitudinal constructs with items making privacy preferences auditable, comparable, and actionable, connecting what laws mandate, what policies promise, and what users value.

\appendix



\begin{ethics}

This work aimed to develop a tool designed to help assess user privacy preferences following ethical considerations, and all procedures followed were IRB-approved. Expert participants (\textit{n = }54) provided informed consent, could withdraw without penalty, and none of our procedures induced harm or placed experts at risk. No conflicts of interest exist with any experts. Expert identities are anonymized, and post-acceptance we will explicitly acknowledge experts that chose to be acknowledged.
\end{ethics}

\begin{openscience}

Upon acceptance, the complete item bank (kept and rejected statements) with expert annotations and justifications, code and RAG evaluation/clustering pipeline will be released. For transparency under review, we provide a subset of kept and rejected statements with expert justifications, and the prompt for complementary LLM validation in https://anonymous.4open.science/r/GPPI/, with functionality as described in \textsection\ref{sec:accessSection}. 

\end{openscience}

\begin{ai}

All decisions were human-made, and instrument and annotations were also human-made. No AI-generated text appears in the paper or the developed instrument. OpenAI o3 was used for complementary validation of expert review (\textsection\ref{sec:LLM}). ChatGPT and Claude were used for pre-submission grammar and spelling checks, with grammatical and spelling fixes applied manually when required.
\end{ai}


\section{Expert Panel (Stage 2 and 3)}
\label{app:experts}
\noindent Anonymized expert panel used for statement validation (Stage 2) and for thematic structure validation (Stage 3). 

\begin{table}[h]
\caption{Expert Panel Composition (Stage 2 and 3, $n = 4$).}
\label{tab:validation_experts}
\centering
\begin{tabular}{@{}lccc@{}}
\toprule
ID & Expertise Domain & Years & Relevant Pubs \\
\midrule
A & Security \& Privacy & 12 &  18 \\
B & Security \& Privacy & 8 & 8 \\
C & HCI/Privacy & 15 & 9 \\
D & HCI/Privacy & 10 & 11 \\
\bottomrule
\end{tabular}
\end{table}

\FloatBarrier 

\section{Additional Clustering Validation Details}
\label{app:ablation}
\noindent To verify that each feature modality contributes to cluster quality, we ran an ablation across six configurations: statement embeddings, keyword embeddings, article-context embeddings, all pairwise concatenations, and the full three-way concatenation. We report the Silhouette score (higher is better) computed on the validation split for each configuration.

\begin{table}[h]
\centering
\small
\caption{Ablation study: cluster quality across feature combinations. Scores are means over 5 random seeds (std $\leq 0.01$), cosine distance, computed in the same space used for clustering.}
\label{tab:ablation}
\begin{tabular}{lcc}
\toprule
\textbf{Feature Configuration} & \textbf{Dimensions} & \textbf{Silhouette Score} \\
\midrule
Statement embeddings only & 384  & 0.23 \\
Keyword embeddings only   & 384  & 0.15 \\
Article context embeddings only & 384 & 0.20 \\
Statement + Keywords      & 768  & 0.29 \\
Statement + Article context & 768 & 0.32 \\
Keywords + Article context & 768 & 0.27 \\
\textbf{All three (final approach)} & \textbf{1,152} & \textbf{0.39} \\
\bottomrule
\end{tabular}
\end{table}

\noindent We selected $k \in \{5,\ldots,15\}$ using Silhouette (higher is better), Davies-Bouldin (lower is better), and Calinski-Harabasz (higher is better), averaging over 5 random seeds (std $\leq 0.01$). Results support $k{=}10$: Silhouette peaked at \texttt{0.392} (k=9: \texttt{0.384}, k=11: \texttt{0.387}); Davies-Bouldin reached a local minimum of \texttt{1.26} (k=9: \texttt{1.31}, k=11: \texttt{1.28}); and Calinski-Harabasz showed only marginal gains beyond $k{=}9$ (\texttt{531} at $k{=}9$ vs.\ \texttt{548} at $k{=}10$, \texttt{+3.2\%}; \texttt{551} at $k{=}11$, \texttt{+0.5\%} over $k{=}10$). We fixed $k{=}10$ to align clusters with interpretable GDPR parent themes; given overlapping semantics across themes, Silhouette values in the $0.3$-$0.4$ range are expected for short-text clustering.

\FloatBarrier

\section{Statement Processing Examples}
\label{app:sample}

Table~\ref{tab:statement-examples} traces three sample statements through all pipeline stages, illustrating the range of outcomes: a statement retained without revision (Example~1), a statement retained after Stage~2 revision (Example~2), and a statement removed at Stage~4 for construct misalignment (Example~3).

\begin{table*}[t]
\centering
\caption{Three representative statements traced through all pipeline stages: retained without revision (Example 1), retained after Stage 2 revision for incomplete representation (Example 2), and removed at Stage 4 (Example 3).}
\label{tab:statement-examples}
\small
\renewcommand{\arraystretch}{1.4}
\begin{tabular}{@{} p{2.4cm} p{4.4cm} p{4.4cm} p{4.4cm} @{}}
\toprule
& \textbf{Example 1: Retained} \newline (no revision)
& \textbf{Example 2: Retained} \newline (revised at Stage 2)
& \textbf{Example 3: Removed} \newline (Stage 4 consensus) \\
\midrule

\textbf{GDPR Source}
& Art.~20(4) \newline Ch.~3: Rights of the Data Subject
& Art.~7(3) \newline Ch.~2: Principles
& Art.~68(3) \newline Ch.~7: Cooperation and Consistency \\
\addlinespace[3pt]

\textbf{Original Provision}
& \emph{``The exercise of the right referred to in paragraph 1 of this Article shall be without prejudice to Article 17. That right shall not apply to processing necessary for the performance of a task carried out in the public interest or in the exercise of official authority vested in the controller.''}
& \emph{``The data subject shall have the right to withdraw his or her consent at any time. The withdrawal of consent shall not affect the lawfulness of processing based on consent before its withdrawal. Prior to giving consent, the data subject shall be informed thereof. It shall be as easy to withdraw as to give consent.''}
& \emph{``The Board shall be composed of the head of one supervisory authority of each Member State and of the European Data Protection Supervisor, or their respective representatives.''} \\
\addlinespace[3pt]

\textbf{Stage 1: Statement}
& ``I would like my personal data to be processed only if my right to transfer my data to another organization does not affect my right to have my data erased when applicable. I also acknowledge that this right to transfer my data might not apply when my data is needed for tasks carried out in the public interest or for official duties.''
& ``I would like my personal data to be processed only if I can withdraw my consent at any time and withdrawal is as easy as giving consent.''
& ``I would like my personal data to be processed only if the European Data Protection Board includes the head of one national data protection authority from each EU Member State, along with the European Data Protection Supervisor (or their designated representative), so that every country and the EU-level overseer have a voice in decisions.'' \\
\addlinespace[3pt]

\textbf{Stage 2: Expert Review}
& Round~1: 4/4 Agree. \newline Round~2: 4/4 Agree. \newline No revision needed.
& Round~1: 3/4 Agree; 1~expert flagged \emph{incomplete representation}: the statement omitted that withdrawal does not affect lawfulness of prior processing, and lacked the ``without detriment'' safeguard. \textit{Revised to}: ``\ldots only if my consent can be withdrawn and I am clearly informed of my right to withdraw consent at any time, and if withdrawing consent is just as easy as giving it, without affecting the lawfulness of past processing.'' \newline Round~2: 4/4 Agree.
& Round~1: 4/4 Agree. \newline Round~2: 4/4 Agree. \newline No revision needed. \\
\addlinespace[3pt]

\textbf{Stage 3: Theme}
& Parent: \emph{Data Subject Transparency and Rights}. \newline Subtheme: \emph{Right to Data Portability}.
& Parent: \emph{Lawful Processing Foundations}. \newline Subtheme: \emph{Conditions for Valid Consent}.
& Parent: \emph{European Data Protection Board}. \newline Subtheme: \emph{Tasks of the EDPB}. \\
\addlinespace[3pt]

\textbf{Stage 4: Consensus Review}
& 5/5 Keep. \newline \textcolor{pastelgreen}{\textbf{Retained.}}
& 5/5 Keep. \newline \textcolor{pastelgreen}{\textbf{Retained.}}
& 0/5 Keep. \newline \textcolor{pastelred}{\textbf{Removed.}} \newline Sample Reason: ``This describes the composition of an inter-authority governance body, not a privacy preference or choice that a subject can exercise.'' \\

\bottomrule
\end{tabular}
\end{table*}

\section{Complementary LLM Cluster Coherence Validation (Details).}
\label{appendix:llmdetails}
The following details the complementary LLM validation described in \textsection~\ref{sec:LLM}.

We used few-shot learning~\cite{brown2020fewshot,min-etal-2022-rethinking} to validate the consistency and reliability of our expert labels as a complement to human expert review. We included one-third of the labeled statements in a prompt, and instructed an LLM to create sub-theme and parent theme labels for the remaining statements. Each few-shot example consisted of the statement text paired with its assigned parent theme and subtheme with no additional metadata provided to the LLM. When a new parent theme or subtheme is created, it gets appended to a list. The LLM refers to the list when evaluating any new statement, reusing labels with a high match to avoid recreating parent themes and subthemes that have extensive overlap. The few-shot validation was built with LangChain~\cite{langchain_github}, integrating chain-of-thought (CoT) reasoning model (OpenAI \texttt{o3-2025-04-16}~\cite{openai2025o3}) with a custom retrieval augmented generation (RAG)~\cite{lewis2020RAG, gao2023retrieval} using FAISS vector retrieval~\cite{johnson2017FAISS, douze2025faiss}. We used established practices for reducing hallucinations in domain-specific tasks~\cite{lewis2020RAG,gao2023retrieval} by providing one-third of manually classified statements and official GDPR text rather than pre-trained knowledge. Our approach leverages LLMs' effectiveness in qualitative analysis~\cite{bijker2024chatgpt, tai2024examination}, showing that carefully prompted LLM ``judges'' can assess semantic coherence with strong correspondence to human ratings on open-ended technical responses~\cite{liu2023g, zheng2023judging}.

We manually assessed and computed cosine similarity between LLM- and human-assigned parent themes and subthemes using sentence-transformer embeddings~\cite{reimers2019sentencebertsentenceembeddingsusing, cer2012018universal} (all-MiniLM-L6-v2~\cite{allminilm6v2}). A 3-fold cross-validation achieved 85\% accuracy, a benchmark commonly used for semantic equivalence~\cite{tumre-etal-2025-improved, li-2024-tracing, viggiato2022identifying}. Scores (mean: 92.3\%, range: 85.2 to 96.8\%) were calculated as the percentage of statement pairs where human and LLM classifications matched. The remaining mismatches (15\%) are expected in few-shot settings: some validation statements address GDPR concepts underrepresented in the one-third seed set, producing lower similarity scores. Given Stage 2's 100\% expert agreement on regulatory accuracy and Stage 3's expert confirmation of thematic structure, these scores reflect coverage gaps in the seed sample rather than thematic incoherence.
The LLM flagged no misplaced statements, confirming consistency of clustering and LLM interpretability required for downstream integration with automated policy annotation pipelines (\textsection\ref{sec:taxandautomate}). This triangulation~\cite{denzin2017research,flick2018triangulation} provides validity beyond single-method validation~\cite{campbell1959convergent}, with each method addressing a distinct threat: expert review guards against regulatory inaccuracy, while LLM classification provides an independent consistency check free of conformity effects in human panels.

\end{document}